\documentclass[a4paper,11pt]{article}
\usepackage{url}
\usepackage{jinstpub} % for details on the use of the package, please see the JINST-author-manual
\usepackage{lineno}
\usepackage{multirow}
\usepackage[version=4]{mhchem}
\title{\boldmath Electromagnetic energy calibration of the SoLid detector with horizontal muons}

\author[b]{Y. Abreu}
\author[j]{Y. Amhis}
\author[c]{L. Arnold}
\author[h]{G. Barber}
\author[b]{W. Beaumont} 
\author[a]{S. Binet}
\author[p]{I. Bolognino}
\author[j]{M. Bongrand} 
\author[h]{J. Borg}
\author[j]{D. Boursette} 
\author[e]{V. Buridon}
\author[k]{B. C. Castle} 
\author[a]{H. Chanal} 
\author[c]{K. Clark}
\author[l]{B. Coupé}
\author[a]{P. Crochet}
\author[c]{D. Cussans}
\author[d]{J. D’Hondt}
\author[e]{D. Durand}
\author[m]{T. Durkin}
\author[i]{M. Fallot}
\author[e]{D. Galbinski}
\author[o]{S. Gallego}
\author[l]{L. Ghys}
\author[i]{L. Giot}
\author[h]{K. Graves}
\author[e]{B. Guillon} 
\author[q]{S. Hayashida}
\author[i]{D. Henaff}
\author[h]{B. Hosseini} 
\author[j]{S. Jenzer}
\author[l]{S. Kalcheva}
\author[d]{L.N. Kalousis}
\author[d]{R. Keloth}
\author[k,o]{L. Koch}
\author[g]{M. Labare}
\author[e]{G. Lehaut} 
\author[c]{S. Manley}
\author[j]{L. Manzanillas}
\author[l]{J. Mermans}
\author[g]{I. Michiels} 
\author[a]{S. Monteil}
\author[g,l]{C. Moortgat}  
\author[c,m]{D. Newbold} 
\author[e]{V. Pestel}
\author[c]{K. Petridis}
\author[b]{I. Piñera}
\author[l]{L. Popescu}
\author[b,f]{A. De Roeck} 
\author[j]{N. Roy}
\author[g]{D. Ryckbosch} 
\author[k]{N. Ryder}
\author[c]{D. Saunders}
\author[j]{M.-H. Schune} 
\author[i]{M. Settimo}
\author[b]{H. Rejeb Sfar}
\author[j,n]{L. Simard}
\author[e]{A. Vacheret}
\author[g]{G. Vandierendonck}
\author[l]{S. Van Dyck}
\author[d]{P. Van Mulders}
\author[b]{N. van Remortel}
\author[b,d]{S. Vercaemer}
\author[r]{M. Verstraeten}
\author[i]{B. Viaud}
\author[s,o]{A. Weber}
\author[a]{M. Yeresko}
\author[i]{F. Yermia}

\affiliation[a]{Université Clermont Auvergne, CNRS/IN2P3, LPCA, Clermont-Ferrand, France}
\affiliation[b]{Universiteit Antwerpen, Antwerpen, Belgium}
\affiliation[c]{University of Bristol, Bristol, United Kingdom}
\affiliation[d]{Vrĳe Universiteit Brussel, Brussel, Belgium}
\affiliation[e]{Normandie Univ, ENSICAEN, UNICAEN, CNRS/IN2P3, LPC Caen, Caen, France}
\affiliation[f]{CERN, 1211 Geneva 23, Switzerland}
\affiliation[g]{Universiteit Gent, Gent, Belgium}
\affiliation[h]{Imperial College London, Department of Physics, London, United Kingdom}
\affiliation[i]{SUBATECH, Nantes Université, IMT Atlantique, CNRS/IN2P3, Nantes, France}
\affiliation[j]{IJCLab, Univ Paris-Sud, CNRS/IN2P3, Université Paris-Saclay, Orsay, France}
\affiliation[k]{University of Oxford, Oxford, United Kingdom}
\affiliation[l]{SCK-CEN, Belgian Nuclear Research Centre, Mol, Belgium}
\affiliation[m]{STFC, Rutherford Appleton Laboratory, Harwell Oxford, and Daresbury Laboratory, Warrington, United Kingdom}
\affiliation[n]{Institut Universitaire de France, Paris, France}
\affiliation[o]{Johannes Gutenberg-Universität Mainz, Mainz, Germany}
\affiliation[p]{Department of Physics, The University of Adelaide, Adelaide, SA 5005, Australia}
\affiliation[q]{Kings College London, London, United Kingdom}
\affiliation[r]{Ecole Royale Militaire/Koninklijke Militaire School, Plasma Physics Laboratory, Brussel, Belgium}
\affiliation[s]{Fermi National Accelerator Laboratory, Batavia, USA}

\emailAdd{herve.chanal@clermont.in2p3.fr, Stephane.Monteil@clermont.in2p3.fr, mykhailo.yeresko@clermont.in2p3.fr}

\abstract{SoLid is a neutrino experiment at very-short baseline searching for active-to-sterile oscillations of reactor antineutrinos. The detection principle is based on the pairing of two types of \textit{solid}\@ scintillators: polyvinyl toluene and $^6$Li:ZnS(Ag), which is a new technology used in this field of Physics. In addition to good neutron-gamma discrimination, this setup allows the detector to be highly segmented; the basic detection unit is a 5~cm cube. High segmentation provides numerous advantages including precise localisation of the Inverse Beta Decay (IBD) products, the derivation of an antineutrino energy estimator based on the isolated positron energy, and a powerful background reduction tool that relies on the topological signature of the signal. Finally, the system is read out by a network of wavelength-shifting fibres coupled to photosensors. A relative electromagnetic calibration is performed with horizontal cosmic muons. This source poses the simplest calibration problem in which a single detection unit is involved. In addition, large muon energy deposits allow us to perform a calibration at the most detailed level (\textit{i.e.}\@ per fibre) and to accurately define the fraction of energy escaping to neighbouring detection cells. A statistical precision at the sub-percent level is reached. The paper also discusses two methods to calibrate the absolute energy scale and presents their implementation and results. The first method relies on horizontal muons, though the precision is limited to around 10\% because of the uncertainty in the energy distribution of such muons. A novel, alternative method based on the radioactive americium-beryllium source is proposed. It takes advantage of the electron-positron pair-production process and provides a calibration point at 3.4~MeV (\textit{i.e.}\@ in the core of the IBD positron spectrum). The paper is concluded with various cross-check including a determination of the energy spectrum of the standard cosmogenic background candle: $^{12}$B.}


\arxivnumber{YYY.XX} % Only if you have one

\begin{document}
\maketitle
\flushbottom

\section{Introduction}
\label{sec:Introduction}

The SoLid (Search for Oscillations with a Lithium-6 Detector) experiment is located in the vicinity of the Belgian Reactor 2 (BR2) research reactor at the SCK CEN site in Mol, Belgium. The experiment aims to make a precise measurement of the reactor antineutrino flux at a very short baseline (6.3 - 8.9 m) to search for oscillations of neutrinos to a sterile state. The second goal is to study the $^{235}$U energy spectrum with regard to the ``5~MeV bump''~\cite{5mevbumpdoublechooz}. The first measurement provides information on the so-called Reactor Antineutrino Anomaly (RAA)~\cite{Mention_2011} and the Gallium anomaly~\cite{Barinov_2022}. In particular, it is possible to constrain the 3 + 1 model~\cite{Abazajian:2012ys}, which assumes the existence of an additional light sterile neutrino state. The SoLid experiment was designed to probe the best-fit region of the oscillation parameters with sin$^2(2\theta_s$) $\approx$ 0.1 and $\Delta$m$^2_s$ $\approx$ 1~eV$^2$. The description of the design of the SoLid detector is beyond the scope of this article. The interested reader can consult the detailed discussion reported in Ref.~\cite{SoLid:2020cen}. The following paragraph contains the executive summary required for the description of the calibration procedure. 

\begin{figure}[ht!]
    \centering
    \includegraphics[width=1\textwidth]{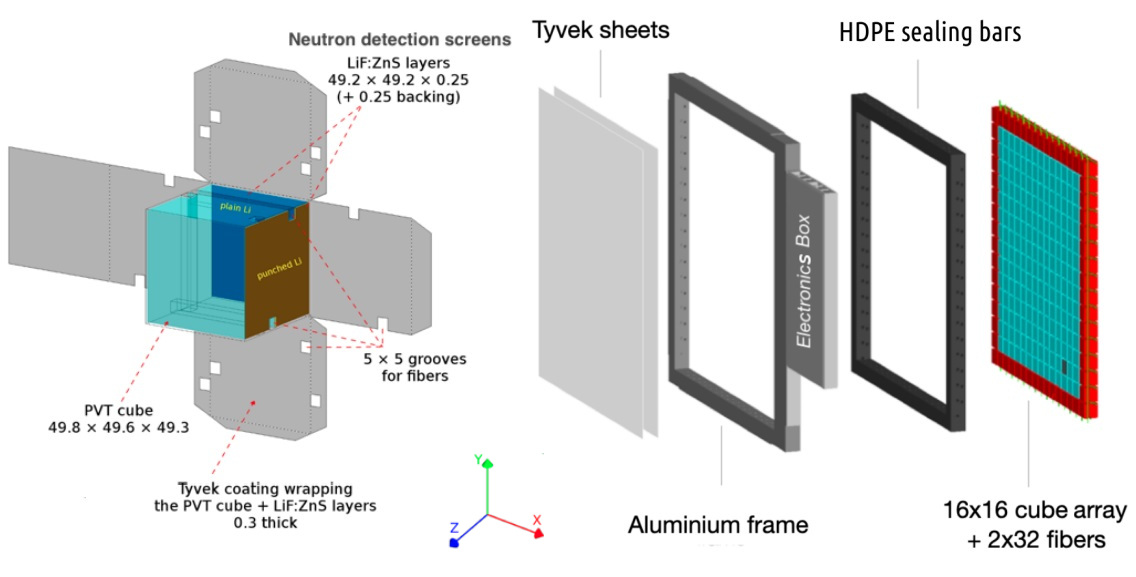}
    \caption{A sketch of the SoLid basic detection unit (left) and detector plane (right).}
    \label{fig:cube_plane_scheme}
\end{figure}

The basic detection unit of the detector is a 5~cm side PolyVinyl Toluene (PVT) cube. PVT is a cheap plastic scintillator with a linear response over a wide range of energy. Each cube has two neutron detection screens (microcomposite $^6$LiF:ZnS(Ag)) placed on adjacent faces. The detection units are combined into planes of 16 $\times$ 16 units each. Each cube is individually wrapped in Tyvek to prevent scintillation light from escaping. Furthermore, each plane is surrounded by two square Tyvek sheets, which optically decouple the planes by further preventing the passage of light between them~\cite{tyvek_data_sheet, SoLid:2020cen}. 
The latter is a very important feature for both calibration and reconstruction, since it simplifies the problem from 3D to 2D. Ten planes make up a module, and the detector comprises five modules in total. The scintillation light from the PVT cube is collected by two vertical and two horizontal WaveLength-Shifting (WLS) fibres that pass through each detection cell. One side of each WLS fibre is covered with a Mylar foil that acts as a mirror to reflect the incoming light. The second side is coupled with Multi-Pixel Photon Counters (MPPCs) that read out the light. The digitised version of the readout is the initial input received from the detector. Figure~\ref{fig:cube_plane_scheme} shows the schematic view of the SoLid detection unit and the detection plane. The origin of the coordinate system corresponds to the position of the reactor core. The $x - y$ coordinates are defined by the orientation of the detector planes, with $x$ being the horizontal direction and $y$ the vertical, respectively. The $z$ axis is perpendicular to the detector planes.

The SoLid detector uses the Inverse Beta Decay (IBD) process ($\bar{\nu}_e + p \to n + e^+$) to detect antineutrinos. The necessary simultaneous detection of the neutron and positron signals justifies the use of two scintillators: neutron detection screens featuring $^6$LiF:ZnS(Ag) and PVT. On average, the IBD neutron receives a kinetic energy of 50~keV, which is much larger than the thermal neutron energy (25~meV). Thus, the neutron is first thermalised via elastic collisions with the nuclei in the PVT. After thermalisation, it is captured on the $^6$Li, which has a very large thermal neutron cross section of 936~b (compared to 0.3~b for hydrogen and 0.5~mb for carbon). The atom breaks into tritium and $\alpha$ particles, generating scintillation photons in the ZnS(Ag) crystals present in the detection screens. The PVT acts not only as a neutron moderator but also as a proton-enriched target for the antineutrino. Both positron and annihilation gamma produce scintillation light in the PVT. This light is subsequently captured by WLS fibres and read out by MPPCs. The high granularity of the detector allows the ionisation and annihilation gamma contributions from the positron to be distinguished. Hence, the antineutrino energy is defined from the IBD process as follows:

\begin{equation}
\begin{gathered}
    E_{\bar{\nu}} + m_p = E_{e^+} + m_{e^+} + E_n + m_n\\
    E_{\bar{\nu}} = E_{e^+} + m_{e^+} + m_n - m_p \approx E_{e^+} + 1.806 \: \textnormal{MeV} \; ,
\end{gathered}
\label{eq:e_estimator}
\end{equation}

\noindent where $m_p$ and $m_n$ denote the masses of the scattered proton and the outgoing neutron, respectively, and $E_i$ with $i \in {e^+,n, \bar{\nu}}$ denotes the kinetic energies of the corresponding particles. Neglecting $E_n$ in the second equation is justified, since the neutron kinetic energies do not exceed 50 keV, much lower than the $\mathcal{O}$(3)~MeV antineutrino energy. Therefore, the antineutrino energy estimator relies on the measurement of the actual energy deposited by the positron, in contrast to the total prompt energy of the event in the case of liquid scintillators. The size of the cube in the SoLid geometry corresponds to the maximum path length of a 10~MeV positron. According to \textsc{Geant4} (version 10.6.0)~\cite{GEANT4:2002zbu} studies, the positron, in fact, deposits its energy in a single cube in 80\% of events. Furthermore, the cube in which the annihilation occurred (Annihilation Cube or AC) is the most energetic cube of the event when the positron energy is above 1~MeV. Thus, an accurate reconstruction of the AC and an accurate calibration of its energy are key to a precise measurement of the antineutrino spectrum; which in turn is of the utmost importance for both oscillation analysis and the ``5~MeV bump'' exploration.

\begin{figure}[ht!]
    \centering
    \includegraphics[width=0.5\textwidth]{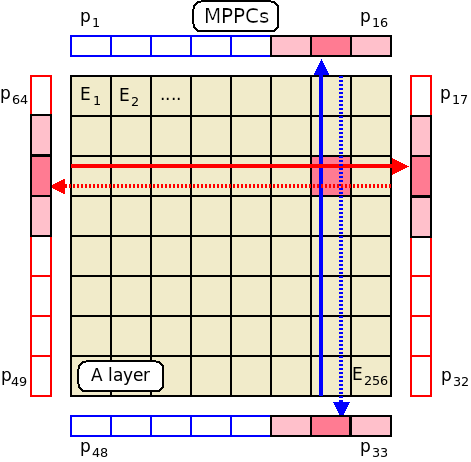}
    \caption{A sketch of the energy deposit (pink cube) in the SoLid detector plane with fired horizontal (red) and vertical (blue) fibres and impacted (pink and light pink) MPPC.}
    \label{fig:ccube}
\end{figure}

However, the detector does not directly deliver the positron energy information at the cube level. The starting point for any analysis is the digitised readout from the MPPCs. Therefore, signal candidates must be defined from this input. More precisely, the MPPC response must be transformed back into the list of detection cubes involved in the event. This list of cubes contains the energy deposits from the annihilation gamma and positron if the mean path length of the annihilation gamma is enough to escape the AC. In addition to verifying the energy estimator suggested in Equation (\ref{eq:e_estimator}), the reconstructed annihilation gamma clusters also provide a very powerful background rejection tool. Figure~\ref{fig:ccube} sketches the basics of the reconstruction problem. If there is an energy deposit $E_j$, then the WLS fibres act as linear projectors to transfer light to the MPPC. As mentioned above, the planes are optically decoupled; hence, the reconstruction problem can be posed for each plane individually. We postulate $f_{ij}$ as the projector from cube \textit{j}\@ to MPPC \textit{i}\@ (generally speaking, $f_{ij}$ represents the fraction of light generated in the given cube received by an individual fibre). The definition of $f_{ij}$ involves the absolute energy scale and the MPPC gain. It is discussed in more detail in Section~\ref{sec:relative}. The readout value for this particular MPPC is calculated as follows:

\begin{equation}
    f_{i,1}\cdot E_{1} + f_{i,2}\cdot E_2 \;+\; ... \;+\; f_{i,256}\cdot E_{256} \;=\; p_i \; ,
    \label{eq:ccube}
\end{equation}

\noindent where $p_i$ corresponds to one of the 64 readout projection values (twice the sum of the number of cubes in rows and columns) and 256 corresponds to the number of cubes in the plane. A similar equation can be written for each individual MPPC. Afterwards, the system of 64 equations can be represented in matrix form as: 

\begin{equation}
      A_{64 \times 256} \times \textbf{E} = \textbf{p} \; ,
\label{eq:ccube_matrix}
\end{equation}

\noindent where $\textbf{p}$ is the column vector of the readout projections. $\textbf{E}$ is a column vector of unknown energy deposits that are determined by the reconstruction procedure. Finally, $A_{64 \times 256}$ is called the system matrix and embodies the best of our knowledge about the detector behaviour at each stage, from light generation to digitisation. Finally, it must be complemented with the absolute energy scale to transform the energy of the cubes from the Analogue-to-Digital Converter units (ADCs) to the physics units (MeV). The description of the methods for solving Equation~(\ref{eq:ccube_matrix}) is beyond the scope of this article. The tools available on the market, together with the baseline choice made for the SoLid experiment, known as the CCube algorithm, are exhaustively discussed in Ref.~\cite{reco_note}. The equation is solved under the assumption that the system matrix is determined elsewhere. The derivation of this matrix and the determination of the absolute energy scale are the subject of this article. 
\section{Cosmic muons as the calibration source}
\label{sec:strategy}

\begin{figure}[ht!]
    \centering
    \includegraphics[width=0.95\textwidth]{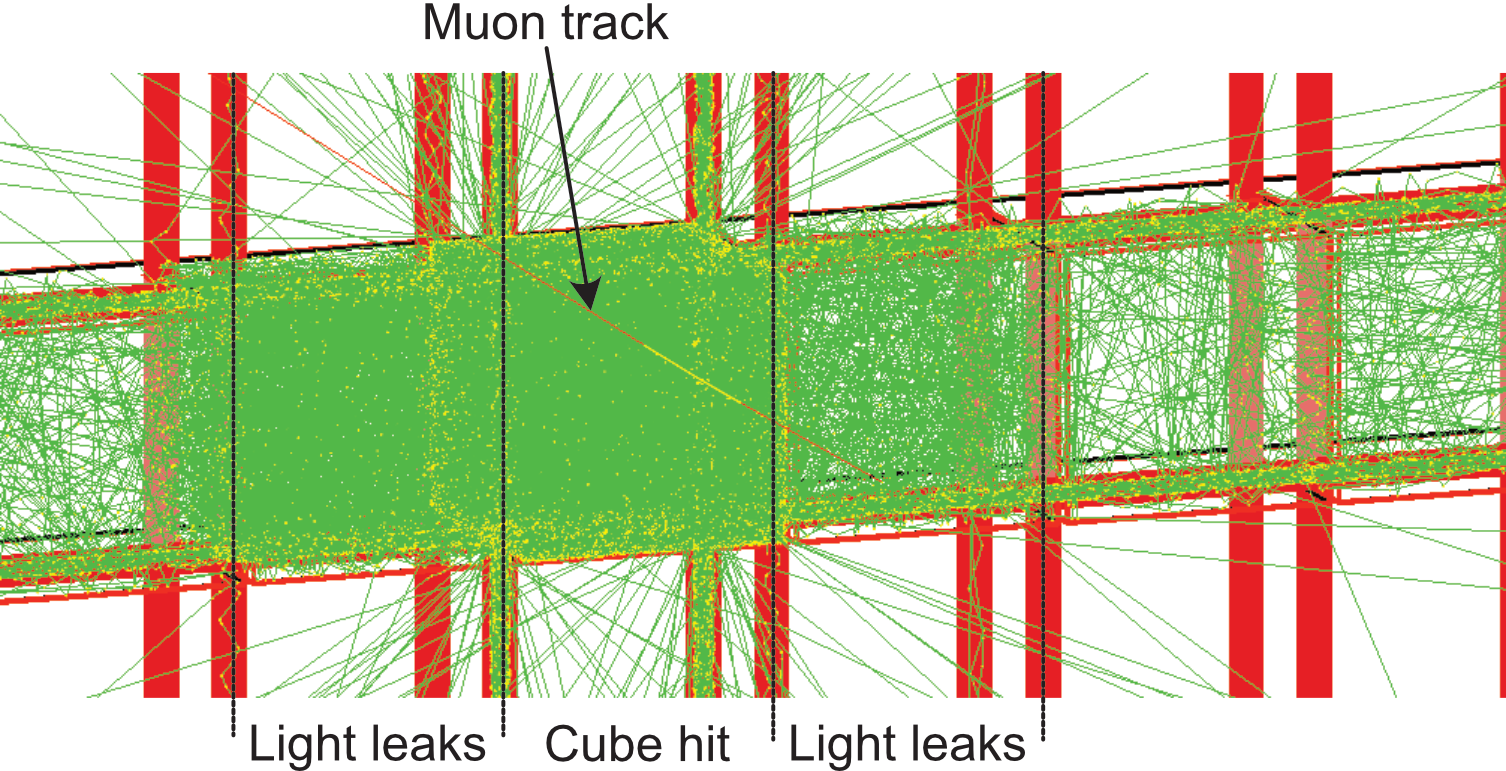}
    \caption{An optical simulation setup with a single muon track launched through the central cube.}
    \label{fig:muon_optical_simulation}
\end{figure}

The projector values $f_{ij}$ in Equation~(\ref{eq:ccube}) are not constrained \textit{a priori}. Their initial features are obtained from a simplified optical simulation. The simulation is performed with the GODDeSS extension framework~\cite{Dietz-Laursonn_2017} for \textsc{Geant4}. It consists of one row of 16 identical PVT cubes without neutron detection screens, but wrapped in reflective Tyvek paper. Both horizontal and vertical WLS fibres are paired with MPPCs, which read out the number of arriving scintillation photons. A single muon is generated such that it crosses the central cube of the system. The setup is sketched in Figure~\ref{fig:muon_optical_simulation}.

First, the simulation shows that the scintillation light is not contained within a single cube. A significant fraction of the photons ($\sim$ 10\%) are propagated to the neighbouring cubes, in particular through the holes created by the WLS positioning tolerances. This effect will be referred to as Light Leaks (LL) in the following. The fraction of photons that end up two cubes away from the crossed cube is at a subpercent level, and thus neglected. Therefore, each energy deposit has only 12 non-zero projector values associated with it. Four of them are related to the fibres in the \textit{main}\@ cube, where the energy deposit has occurred, and another eight are related to the four neighbouring cubes (2 on the horizontal axis and 2 on the vertical) created by the LL. Secondly, the simulation reveals asymmetries in light sharing between the fibres. These asymmetries are observed between the horizontal and vertical fibres. Moreover, the amount of light shared between the two horizontal (vertical) fibres that cross the same cube is not equivalent either. Thus, a per-fibre determination of the System Matrix elements, \textit{i.e.}\@ calibration, is desirable. The final stage of the simplified optical simulation scrutiny is related to the impact of the WLS groove size. The width of the groove is 5~mm, while the diameter of the fibre is 3~mm. In the modified version of the simulation, the width of the groove is changed to match the size of the WLS fibre, so that there is no air gap between the two. This modification demonstrated a drastic change in the distributions of scintillating light, which, in turn, indicates miscalibrations due to the actual geometry of the detector. This effect also varies from cube to cube. Other effects such as temperature or the detector displacements triggered by the movement of the calibration sources can also influence the detector response. The modelling of the miscalibrations cannot therefore rely solely on physics assumptions such as attenuation length or coupling effect. Since all detection units are identical in the \textsc{Geant4} simulation, the modelling cannot rely on it either. There are two ways to address this problem: develop a much more detailed simulation or alternatively use the data directly. 

In summary, the optical simulation shows that the calibration in the SoLid experiment must be performed on the per-fibre level. In addition, the calibration source must provide a way to determine the LL. Accurate measurement of light-sharing features contributes to a precise reconstruction of the positions of the cubes and their energies. Events can therefore be categorised on the basis of their topological characteristics. These calibration requirements are further complemented by the fact that the detector is a dynamic system; \textit{e.g.}\@ the PVT is exposed to ageing, which decreases the number of generated photons and therefore modifies the energy response during data taking. As such, the calibration procedure must be performed regularly, with a frequency defined by the precision of the method. This in turn provides the third requirement: the precision of the approach has to be at the percent level, which requires enough statistics for all the fibres. Finally, let us return to Equation~(\ref{eq:ccube}). In the case of the unique energy deposit $E_j$, for the projection $p_i$, the equation simplifies as follows:

\begin{equation}
    f_{ij}\cdot E_{j} \;=\; p_i \; ,
    \label{eq:ccube_1deposit}
\end{equation}

\noindent where $p_i$ is the direct output of the detector. Hence, if the total energy deposited in the plane is known, the projector value $f_{ij}$ is the only unknown. Therefore, in such a case, all 12 projector values associated with the cube, where the energy deposit occurs, can be measured directly. Cosmic muons can meet all the requirements listed above. On average, muons deposit 1.6~MeV/cm (which translates to 8~MeV for the SoLid detection unit size of 5~cm). As such, it is possible to accurately measure both the light-sharing properties within the main cube traversed by a muon and in the neighbouring cubes impacted by the LL. Furthermore, the muon track typically crosses several cubes in the detector. Thus, it is possible to use a single track to calibrate multiple cubes. However, to fit the condition of the single energy deposit in the plane, only a subset of the cosmic muon sample is considered for calibration purposes. It consists of parts of the muon tracks that are \textit{horizontal}\@ enough to cross only one cube in a plane.
\section{Selection of horizontal muons}
\label{sec:selection}

Horizontal muon candidates are selected from both reactor-off (ROff) and reactor-on (ROn) samples. The selection itself is implemented in the SoLid Analysis Framework (Saffron2). The main objective of the tool is to cluster the waveforms recorded by detectors that are close in time and space. The waveforms within a cluster are divided into three mutually exclusive categories: muons, nuclear signals, and electromagnetic signals. The latter two are aimed at selecting the signals originating from the IBD neutron (nuclear) and positron together with annihilation gammas (electromagnetic), respectively. The exhaustive description of the Saffron2 software can be found in Ref.~\cite{pestel_thesis}. As for muon clusters, they are searched for in the first place that satisfies a single requirement on the fibre read-out: the presence of high-energy channels with an amplitude greater than 200 ADC counts. If the number of these channels is less than 11, the muon candidate is tagged as a Type 0 muon. In most cases, Type 0 muons correspond to the so-called clipping muons, which cross only a few cubes on the edge of the detector. In other cases, the muon candidate deposits energy in a larger number of cubes that most likely form a track. The latter assumption is checked with two weighted least-squares linear fits (vertical and horizontal projections). 

The candidates are further divided into two categories according to the convergence of the fits: Type 1 if either one of the fits has failed; Type 2 if both fits were successful. The horizontal muons used for the calibration were selected exclusively from the Type 2 muon sample. It is important to mention that the fits are performed solely with the fibre-level information, \textit{i.e.}\@ the reconstruction algorithm is not involved in determining the positions of the cubes on the track. Therefore, one can simultaneously solve Equation~(\ref{eq:ccube_matrix}) and determine the System Matrix. Several additional requirements are applied to Type 2 muons to increase the quality of the sample and ensure that there is indeed a single cube hit in the plane. First, the start and end cubes of the track have to be on the detector border. This requirement allows muons that are captured while crossing the detector (stopping muons) and as such have different stopping power to be rejected. Second, the muon track length must be greater than or equal to 7 cubes. This rejects candidates that are likely excited nuclei. Third, the slopes $x/z$ and $y/z$ must provide $\cos \theta$ less than 0.8, where $\theta$ is the polar angle of the muon track (approximately 40$^{\circ}$). The development of muon selection is discussed in Ref.~\cite{Giel}. 

\begin{figure}[ht!]
    \centering
    \includegraphics[width=1\textwidth]{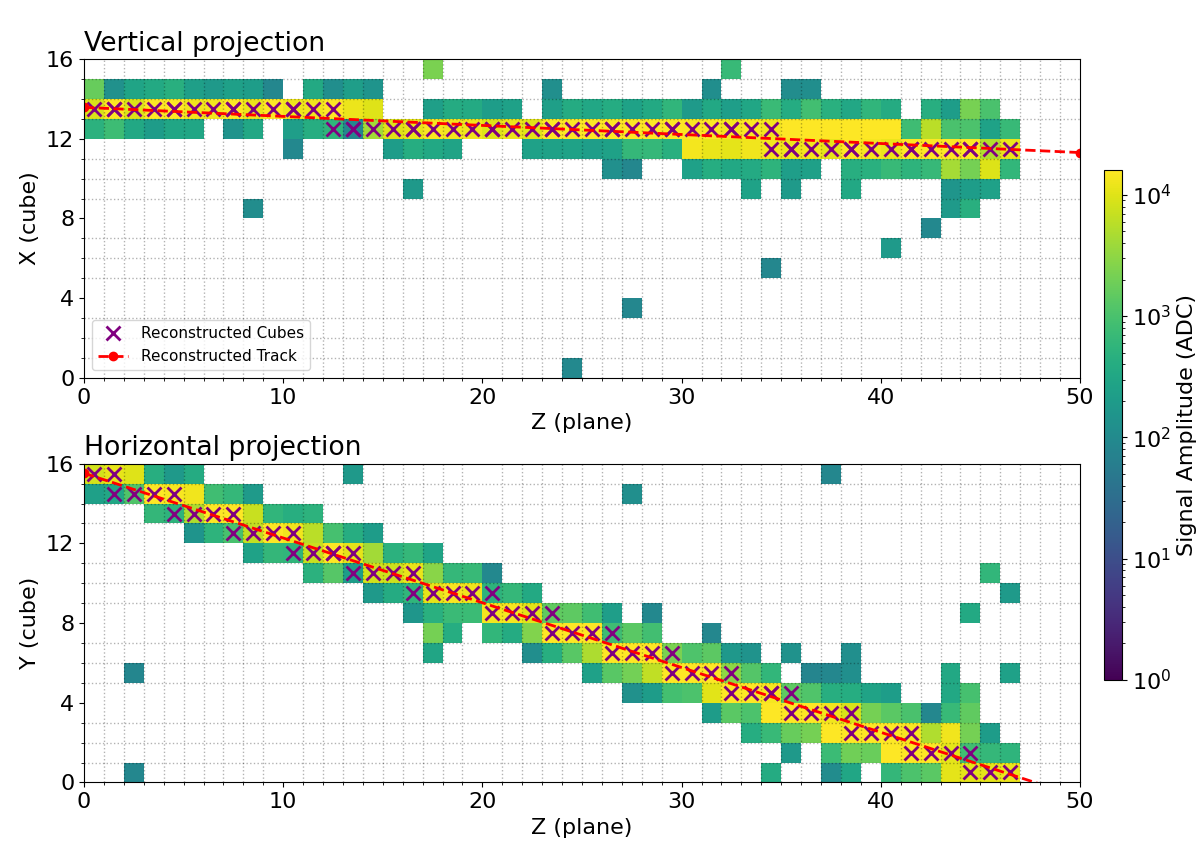}
    \caption{An example of the horizontal muon track obtained from ROff data.}
    \label{fig:horizontal_muon_track}
\end{figure}

\begin{figure}[ht!]
    \centering
    \includegraphics[width=0.55\textwidth]{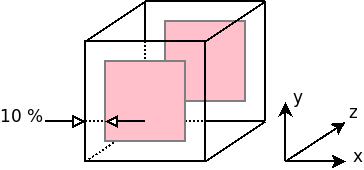}
    \caption{The entrance and exit windows (shown in pink), that the reconstructed muon track must follow.}
    \label{fig:entrance_window}
\end{figure}

An example of a Type 2 muon track that meets all the listed conditions is shown in Figure~\ref{fig:horizontal_muon_track}. However, not all cubes along this track are used in the calibration procedure. To select the planes in which the muon has indeed deposited its energy in a single detection unit, the cube is kept if and only if the reconstructed track enters and exits the cube through a fiducial window. This window is shown in pink in Figure~\ref{fig:entrance_window} and represents the area which is at least 10\% (of the cube size) away from each border. It lifts the ambiguity of a choice of the cube of interest when the track passes close to the border. The set of such cubes from the dedicated muon tracks is referred to as horizontal muons in the following. 

There are 4 fibres traversing each SoLid cube. If there is an energy deposit in the cube, each fibre should receive a fraction of the scintillating photons. However, there are several scenarios for which this is not the case. The first one appears when the energy contribution is not enough for the fibre photons to meet the threshold level; since muons typically deposit large amounts of energy on average, this is not the main case. Another appears when one or more fibres are \textit{dead}\@, \textit{i.e.}\@ that the fibre does not transport the photons to the MPPC. About 1\% of the fibres are permanently dead (most of these dysfunctional fibres are concentrated in Module 5. This module is the farthest from the reactor core, and hence receives the least statistics), or temporarily in dysfunction. 

\begin{figure}[ht!]
    \centering
    \includegraphics[width=1\textwidth]{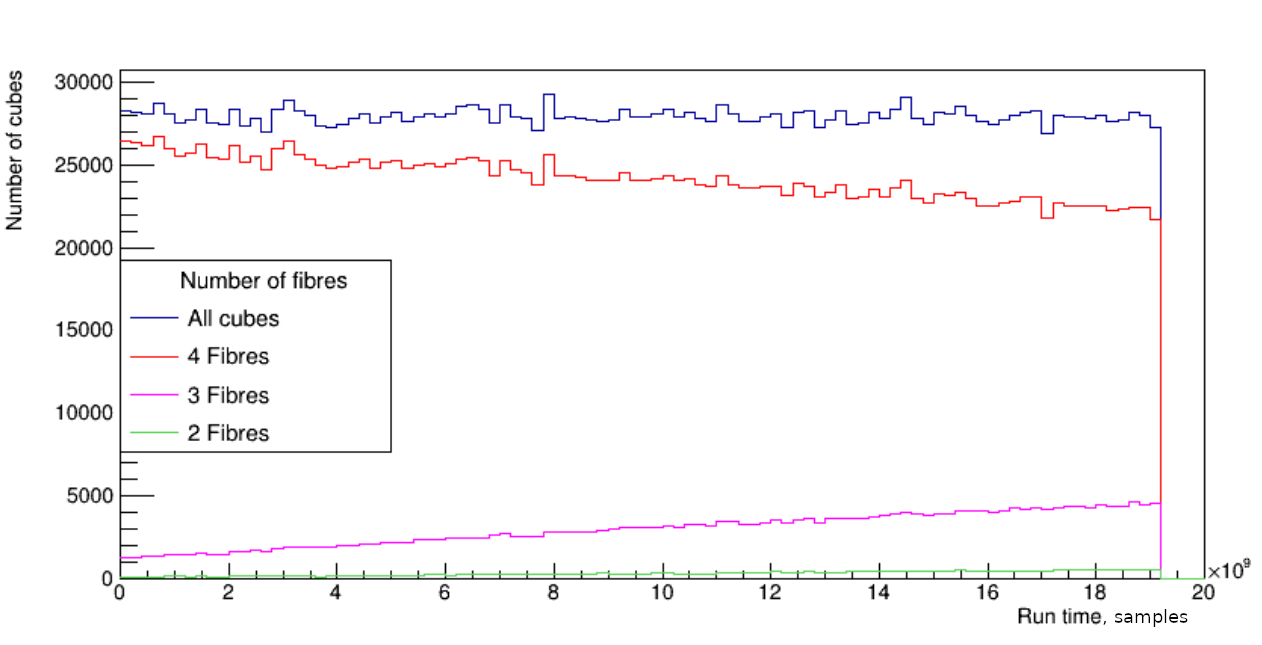}
    \caption{The time evolution of the number of cubes reconstructed with different number of fibres within a run of ROff data. The total number of cubes (blue) is shared among 4 (red), 3 (purple) and 2 (green) fibres.}
    \label{fig:dying_fibres}
\end{figure}

The second scenario occurs if a fibre receives too many scintillating photons (\textit{e.g.}\@ stopping muons) to the extent that it exceeds the processing capacity of the electronics. According to ROff studies, a single channel saturates at 15700 ADC counts. When a channel saturates, the corresponding event buffer is required to be reset in order to transmit again normally the read-out information. This reset occurs only at the beginning of a new physics run that lasts 8 minutes. Therefore, a \textit{saturated}\@ fibre remains dead until the end of the current run. The time evolution of the number of candidates reconstructed with different numbers of fibres is presented in Figure~\ref{fig:dying_fibres}. The \textit{sample} unit of the $x$-axis corresponds to 25~ns. Hence, the figure represents a single physics run of 8 minutes. As a side note, the dying fibre effect also impacts Equation~(\ref{eq:ccube_matrix}). If the fibre dies during the run, the corresponding projector values a$_{ij}$ are switched to zero and restored with the fibre at the beginning of the next run.
\section{Relative calibration}
\label{sec:relative}

As discussed in Section 2, a sample of horizontal muons is considered to obtain the system matrix elements that define a cell-by-cell calibration. For each fibre, projection $p_i$ is calculated by multiplying projector $f_{ij}$ by the energy deposit made in the cube $E_j$. The definition of $f_{ij}$ involves the multiplication of three factors: the MPPC gain ($g_i$), the absolute energy scale ($\beta_j$, which shows the number of photoavalanches generated per unit of energy) and the fraction of light received by the fibre ($a_{ij}$). The result of the multiplication of $E_j$ by $\beta_j$ is the total number of scintillation photons (denoted as $n_{\textnormal{plane}}$) in the given cube, which matches the number of photons in the given plane (since, by construction, horizontal muons hit a unique cube in the plane). The gain for the entire SoLid detector was equalised at the beginning of each data-taking period~\cite{roy_thesis}. As discussed in Section~\ref{sec:absolute}, the absolute scale factors are homogenised for all detection cells. Therefore, it is equivalent (up to a scaling factor) to fill the system matrix with the values of $f_{ij}$ or $a_{ij}$. We arbitrarily chose the latter. In summary, the projection $p_i$ is defined as:

\begin{equation}
    p_i = E_j\cdot f_{ij} = E_j\cdot \beta_j\cdot a_{ij}\cdot g_i = n_{\textnormal{plane}}\cdot a_{ij}\cdot g_i = n_i\cdot g_i = n_i\cdot g \; ,
    \label{eq:relative_calib1}
\end{equation}

\noindent where $n_i$ represents the number of scintillating photons reaching the MPPC $i$. The sum of all projections $p_i$ in the given plane (denoted as $p_{\textnormal{plane}}$) is the only missing piece to compute $a_{ij}$. It is expressed as: 

\begin{equation}
    n_{\textnormal{plane}} = \sum_{i}n_i \implies p_{\textnormal{plane}} = \sum_{i}p_i = \sum_{i}n_i \cdot g = g \cdot \sum_{i}n_i = n_{\textnormal{plane}}\cdot g \; .
    \label{eq:relative_calib2}
\end{equation}

\noindent Finally, the projector $a_{ij}$ is given by:

\begin{equation}
    a_{ij} = \frac{p_i}{p_{\textnormal{plane}}} = \frac{n_i \cdot g}{n_{\textnormal{plane}} \cdot g} = \frac{n_i}{n_{\textnormal{plane}}} \; .
    \label{eq:relative_calib3}
\end{equation}

\begin{figure}[ht!]
    \centering
    \includegraphics[width=0.5\textwidth]{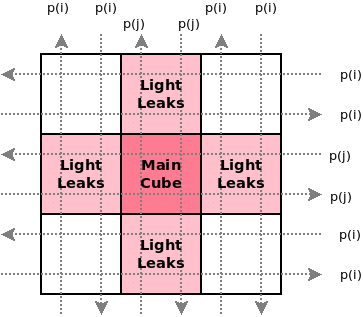}
    \caption{The calibration problem posed by horizontal muon. Scintillation light shared between the hit cube (in pink) with fibre projectors p$_i$ and the cubes affected by the LL (light pink) with projectors p$_j$.}
    \label{fig:calib_problem_init}
\end{figure}

\noindent This formula is equivalent for all fibres (shown in Figure~\ref{fig:calib_problem_init}) of the main cube that is traversed by the muon track, and of the neighbouring cubes that are affected by LL. This formula provides a single value for the projector $a_{ij}$ of a single muon. Once all muons crossing a cube are considered, the distribution of the projector values is obtained. The value $a_{ij}$ that goes into the System Matrix is determined from a fit of a Landau model to the distribution. A transformation that deals with the outliers and ensures a Gaussian behaviour of the distribution is applied, and is described in the next section.

\subsection{KL divergence}
\label{subsec:kl_divegence}

\begin{figure}[ht!]
    \centering
    \includegraphics[width=7cm, height=6cm]{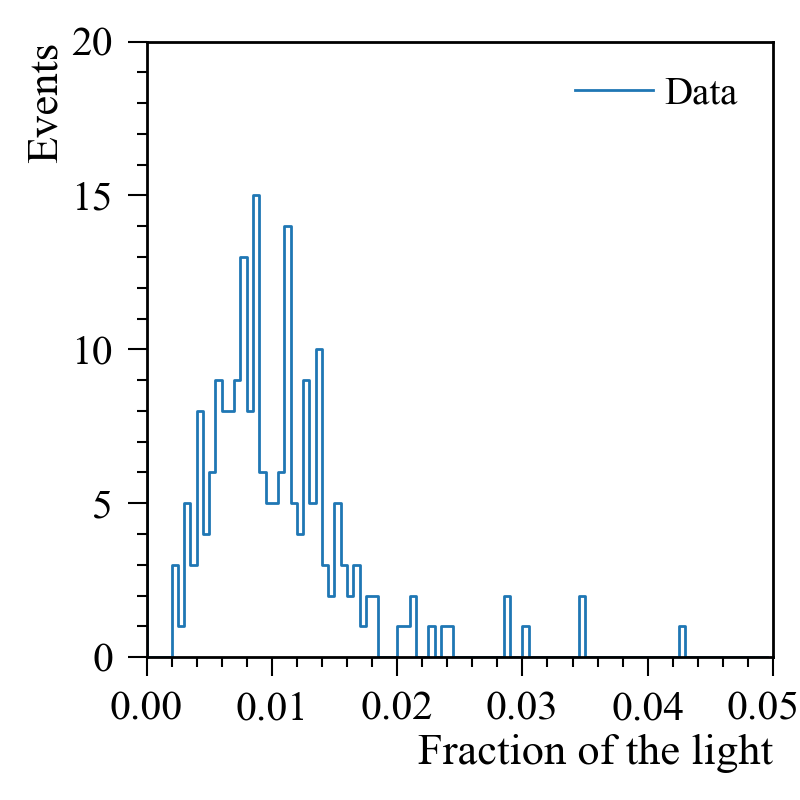}
    \hspace{0.5cm}
    \includegraphics[width=7cm, height=6cm]{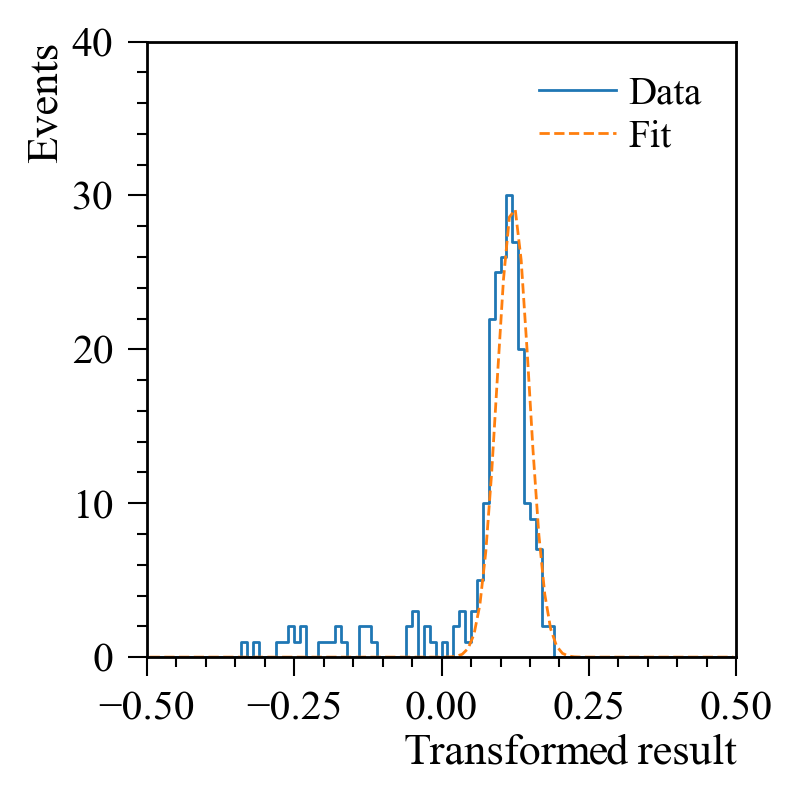}
    \caption{The distributions of the light fraction observed by one of the fibres in the neighbouring cubes (triggered by LL) before (left) and after (right) application of the KL divergence method.}
    \label{fig:kl_strikes}
\end{figure}

The left panel of Figure~\ref{fig:kl_strikes} shows the raw distribution of the $a_{ij}$ values for one of the fibres in the adjacent cube (due to LL). A fit to this distribution with a Landau function would allow us to determine the $a_{ij}$ values. However, the presence of outliers compromises the goodness of fit. There are multiple approaches to address outliers, and the Kullback–Leibler (KL) divergence method~\cite{10.1214/aoms/1177729694} is one of them. The KL divergence method is a type of statistical distance (or dissimilarity test), which quantifies how much a probability distribution $p_1$ is different from a reference probability distribution $p_2$. In the case of the SoLid experiment, both distributions must obey the Poisson law; hence, the KL divergence, slightly modified with respect to Ref.~\cite{10.1214/aoms/1177729694}, is given by:

\begin{equation}
\begin{gathered}
\label{eq:kl_divergence_intermediate}
    m = KL(p_1 \;\|\; p_2) = \sum p_1(x) \cdot \ln{\frac{p_1(x)}{p_2(x)}} = \lambda_1\cdot\ln{\frac{\lambda_1}{\lambda_2}} + (\lambda_1 - \lambda_2) \; .
\end{gathered}
\end{equation}

\noindent This class of equations is solved with a Lambert function $W$~\cite{lambert}. This is a multivalued function with different branches. In our implementation, $\lambda_1$ is either 22.5\% (main cube with four fibres) or 1.25\% (adjacent cubes, 8 fibres). The parameter $m$ is the mean of a Gaussian model fit to the initial distribution of $x$. Equation~(\ref{eq:kl_divergence_intermediate}) is solved for $\lambda_2$ (mean value of the measured distribution) as follows: 

\begin{equation}
    \lambda_2 = \lambda_1 \cdot W[\frac{1}{\lambda_1}\cdot e^{\rho}] \; ,
    \textnormal{where } \rho = 1 + \lambda_1\cdot\ln{\lambda_1} - \frac{m}{\lambda_1} \; .
\end{equation}

\noindent As a result, it is not the distribution of $a_{ij}$ values that is fit. Instead, we fit the statistical distance for each set of projector values from the Poisson law. The outliers in the transformed distribution are much more pronounced and can be straightforwardly excluded from the fit. The result of the application of the KL divergence method is shown in Figure~\ref{fig:kl_strikes}. As such, the total sum of the light-sharing fractions from the four fibres of the main cube and the eight fibres of the adjacent cubes improves from 96\% to 99\%.

\subsection{Homogenisation of the response}

The second stage of the relative calibration is the homogenisation of the response from individual cubes. The characteristics of the detection units (namely the absolute energy scale $\beta_j$) differ, causing a variation in the response of the detector. The energy loss of the muon serves as a calibration reference. The sum of the projections of the 12 fibres involved within the plane is considered as the representation of the energy deposit made by the muon while crossing it (as in the variable $p_{\textnormal{plane}}$ from Equation~(\ref{eq:relative_calib2})). In addition, all muon candidates are selected from Type 2 muons, for which there is a successfully reconstructed track. The path length of the muon within a certain plane is determined from this track information and is further denoted as $x_{\textnormal{track}}$. Thus, the energy loss is estimated as:

\begin{equation}
    \frac{\textnormal{d}E}{\textnormal{d}x} = \frac{p_{\textnormal{plane}}}{x_{\textnormal{track}}} \; .
    \label{eq:de_dx}
\end{equation}

\begin{figure}[ht!]
    \centering
    \includegraphics[width=7cm, height=6cm]{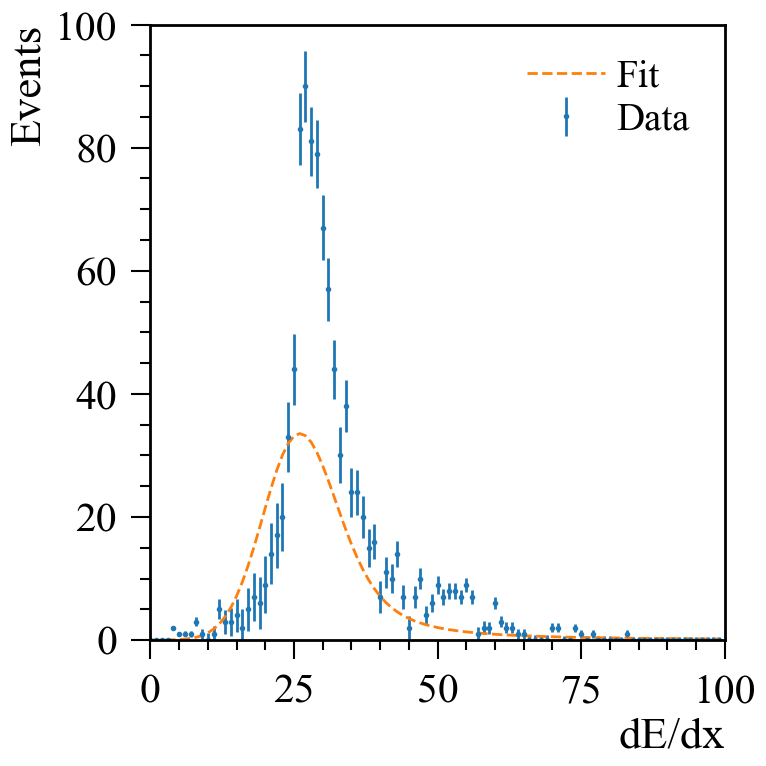}
    \hspace{0.5cm}
    \includegraphics[width=7cm, height=6cm]{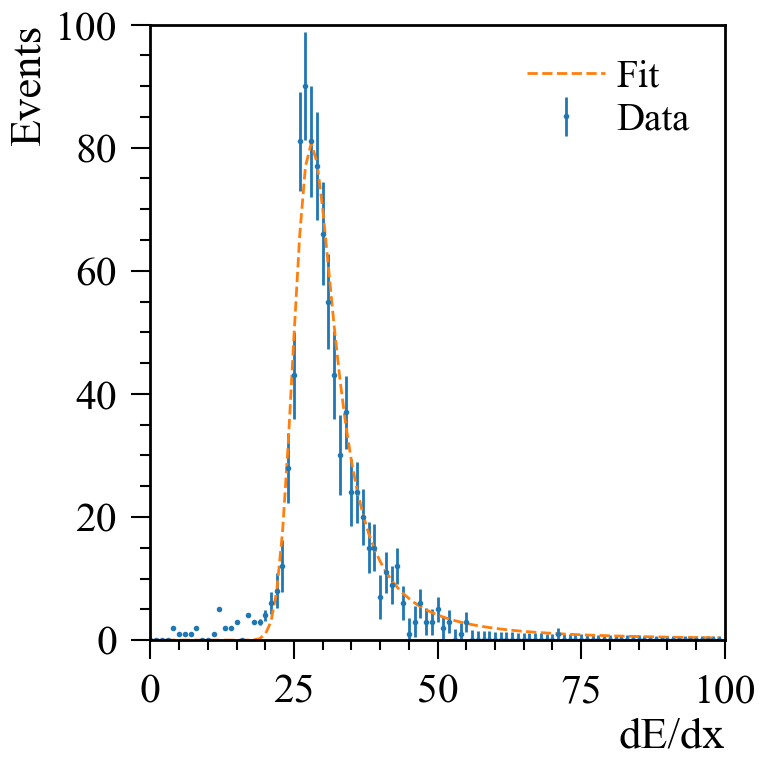}
    \caption{The energy loss distribution before(left) and after (right) application of the 4 active fibre cut.}
    \label{fig:dedx_impr}
\end{figure}

\noindent The energy loss distribution for a given cube is obtained by applying Equation~(\ref{eq:de_dx}) to the subset of muons that cross it. A muon is rejected if the cube is not reconstructed with 4 fibres (except for the cases where the fibre is permanently dead). This additional selection compensates for the dying fibre effect presented in Section~\ref{sec:selection}. The result of the selection is shown in Figure~\ref{fig:dedx_impr}. A Landau function convoluted with a Gaussian is fit to the resulting energy loss distribution. The model has 3 free parameters: the Gaussian width, the Landau scale factor, and its most probable value ($m_{Lj}$, where $j$ shows that the fit result comes from the energy deposit $E_j$). The latter is used as the denominator in the homogenisation procedure.

\begin{equation}
    \label{homogenisation}
    a^{\rm final}_{ij} = a_{ij} \cdot \overline{m}_L \mathbin{/} m_{Lj} \;.
\end{equation}

\begin{figure}[ht!]
    \centering
    \includegraphics[width=0.85\textwidth]{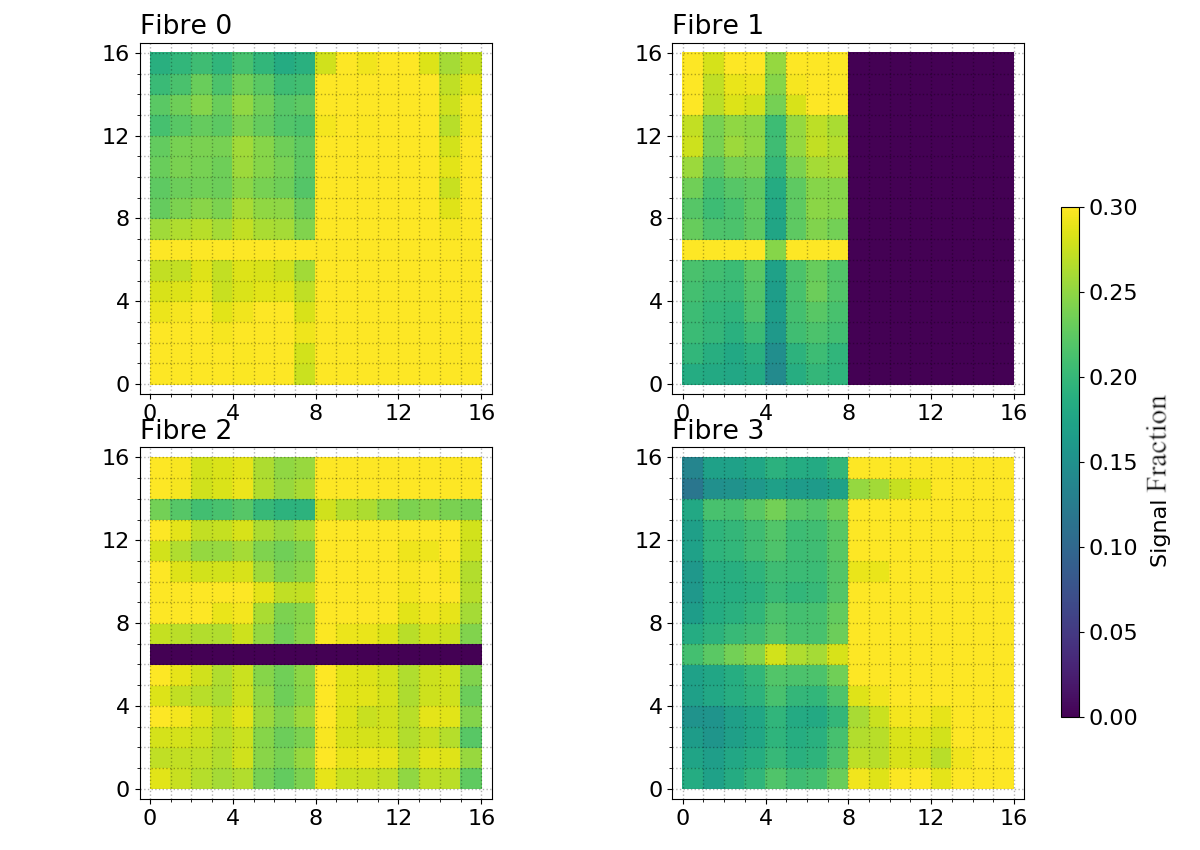}
    \includegraphics[width=0.85\textwidth]{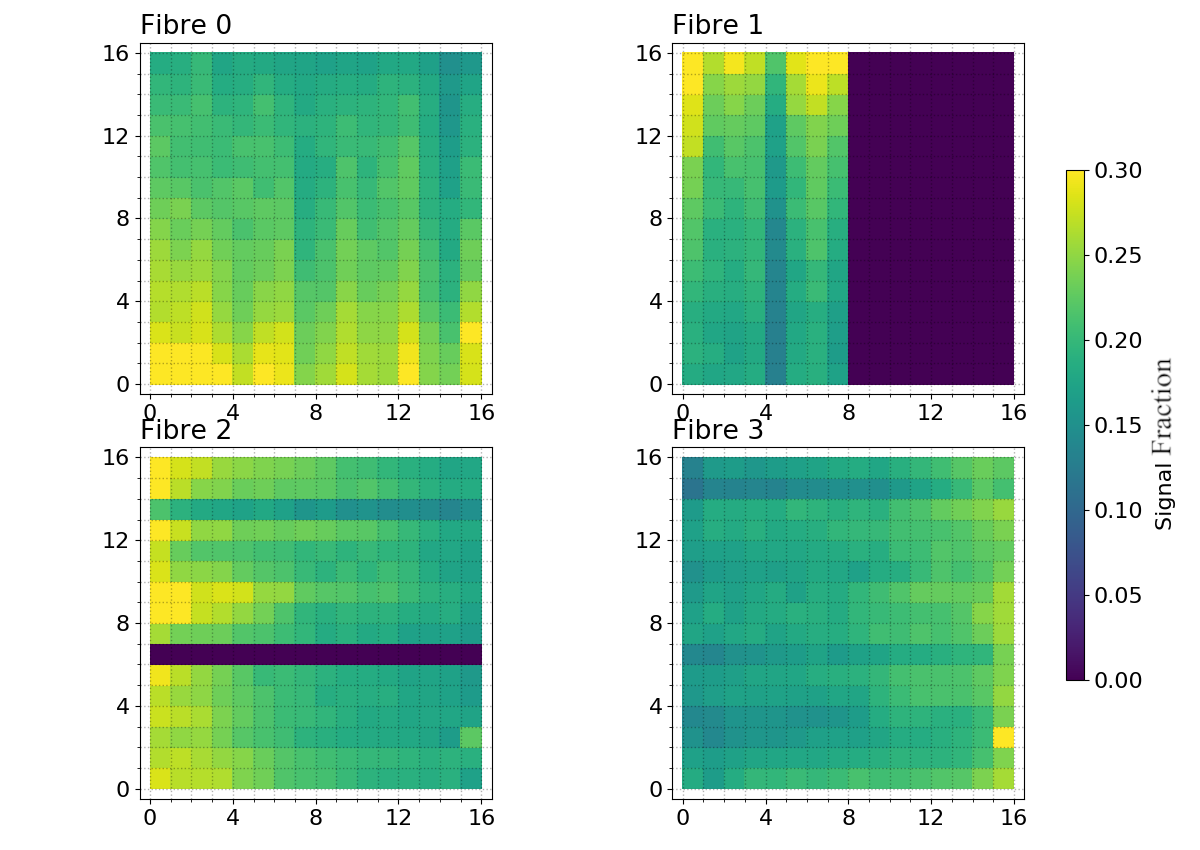}
    \caption{The light sharing distributions ($a_{ij}$ values) for the 4 main fibres of all the cubes in the plane 48 (dark blue corresponds to permanently dead fibres) before (left) and after (right) the homogenising procedure.}
    \label{fig:homogenised_response}
\end{figure}

\noindent The mean detector response is used as the numerator. It is denoted as $\overline{m}_L$ and is calculated as the mean value of the distribution with all the values of $m_{Lj}$ of the individual detection cubes. Projector values $a_{ij}$ are weighted with respect to the average detector response. As a result of the re-weighting, the $\beta_j$ values are aligned for all the detection cells. It concludes the relative calibration of the detector and establishes the final values of the elements of the System Matrix. The impact of the homogenisation procedure is, as expected, the most pronounced for planes with permanently dead fibres. As an illustration, Figure~\ref{fig:homogenised_response} displays the response of the 4 fibres of each main cube in plane 48. The dark blue area (half of the plane) corresponds to the dead fibres. Even under such conditions, the relative calibration works reasonably well and allows the events appearing in that region of the detector to be adequately reconstructed.

\section{Absolute calibration}
\label{sec:absolute}

As a result of the relative calibration, the absolute energy scale is homogenised for all the detection units but remains unknown. The absolute energy scale is a factor (denoted as $\beta_j$ in Equation~(\ref{eq:relative_calib1})) that converts the energy measurement from the units provided by the electronics (ADC) to physical units (MeV). The determination of the (now) universal scale factor $\beta$ completes the electromagnetic calibration of the detector. 

\subsection{Absolute calibration with the horizontal muons}

One possibility of determining the absolute energy scale consists of using the energy-loss distribution from the cosmic horizontal muon. The energy loss values for all the detection cells are gathered in a single distribution to which the Landau function convoluted with the Gaussian is fit. The most probable value of Landau gives the reference value in ADC units \footnote{Technically in pixel avalanches, or PA, but since the gain is equalised for all the fibres, the conversion factor from PA to ADC is known.}. The value in MeV is obtained from a dedicated \textsc{Geant4} simulation. It includes the full-scale detector geometry that is combined with the CRY engine~\cite{cry} to simulate muons. All energy loss values obtained from the simulation are also combined in a single distribution. The ratio between the most probable values extracted from two fits defines the absolute energy scale. 

\begin{figure}[ht!]
    \centering
    \includegraphics[width=0.44\textwidth]{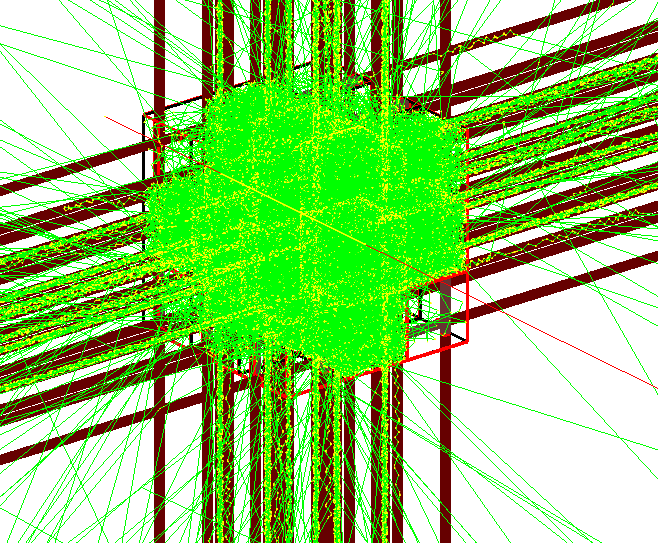}
    \includegraphics[width=0.54\textwidth]{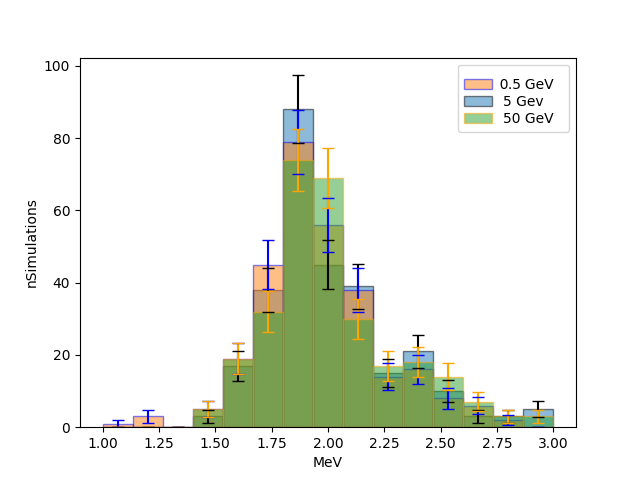}
    \caption{Left: the setup used for the muon energy loss evaluation. Right: the energy loss distributions, obtained for the muons of three different energy regimes. The results are within a broad 10\% agreement.}
    \label{fig:dedx_scrutiny}
\end{figure}

Figure~\ref{fig:dedx_scrutiny} shows the distributions of the energy loss values obtained with muons in three different energy regimes (0.5, 5, and 50~GeV). Variations in the average response are at the level of 10\%. This provides the fundamental limit to the precision of this method. Any improvement would require a more accurate knowledge of the energy distribution of horizontal muons.

\subsection{Absolute calibration with the americium-beryllium source}

A novel method (to our knowledge) has been introduced by making use of an americium-beryllium (AmBe) radioactive calibration source that was used in SoLid to calibrate the neutron signal. The idea consists of measuring the response of the detector to the 4.44~MeV gamma emitted by excited $^{12}$C, with a branching fraction of 60\%. These gammas can experience an electron-positron ($e^+e^-$) pair conversion that provides a well-defined monoenergetic calibration source of 3.4 MeV. Although Compton scattering is the dominant process, the conversion appears in 3.5\% of the cases according to a generic \textsc{Geant4} simulation (with the \textsc{G4EmStandardPhysics} physics list). The presence of the positron makes the calibration signature identical to that of the IBD signal candidates. Therefore, all tools developed for the topological identification of IBD events are highly relevant and can be used to select $e^+e^-$. 

The chain that produces 4.4 MeV gamma rays is as follows:  

\begin{equation}
\begin{gathered}
\label{eq:ambe}
    \ce{^{241}Am -> {\alpha} + ^{237}Np} \\
    \ce{{\alpha} + ^9Be -> {n} + ^{12}C^*} \\
    \ce{^{12}C^* -> {^{12}C} + {\gamma} (4.438 \; \textnormal{MeV}) } .
\end{gathered}
\end{equation}

\noindent Furthermore, neutrons issued from the capture of $\alpha$ in $^{9}$Be, can interact with $^{12}$C nuclei in the detector materials. These neutrons are energetic enough to excite the $^{12}$C nucleus. During further deexcitation, the same 4.44~MeV gamma can emerge as in the nominal decay chain. The initial sample of the calibration signal is therefore increased by 15\%.

\begin{figure}[ht!]
    \centering
    \includegraphics[width=1\textwidth]{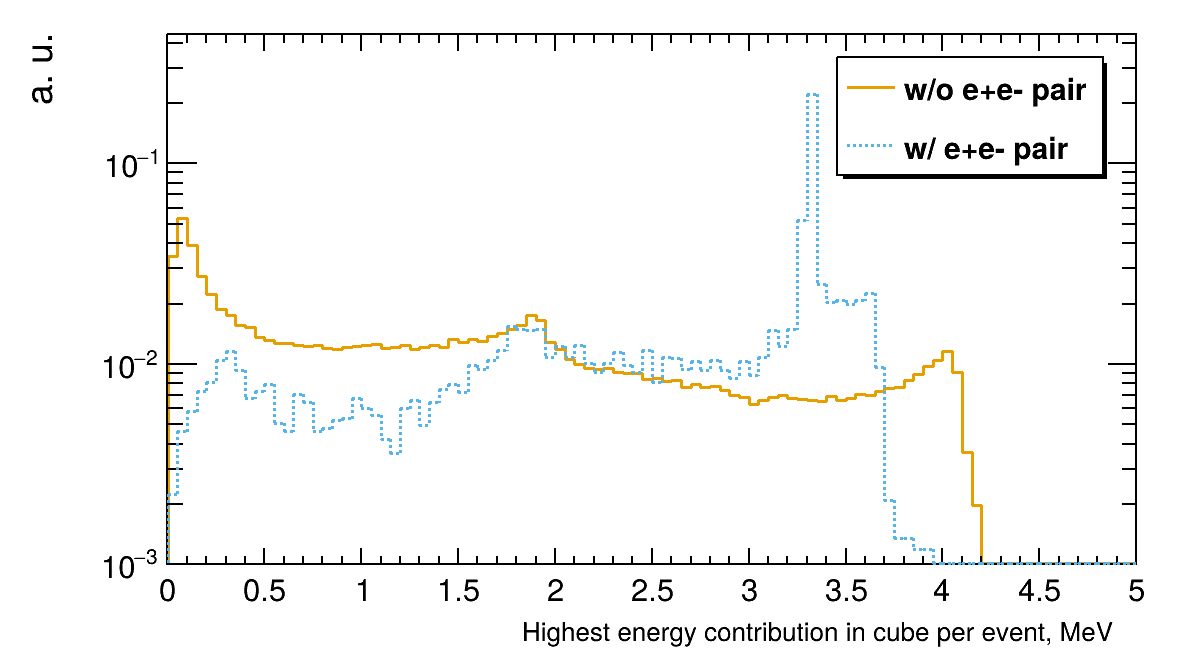}
    \caption{The normalised energy distribution of the cubes with the largest energy deposit in the events from the AmBe \textsc{Geant4} simulation with (dashed blue) and without (solid orange) $e^+e^-$ pair conversion.}
    \label{fig:ambe_es_max}
\end{figure}

The initial information on the electromagnetic part of the signal provided by AmBe is obtained from a dedicated \textsc{Geant4} simulation. It consists of a simulation of the full-scale SoLid detector geometry, where the radioactive source is placed in one of the positions available between the modules. To optimise the computational cost of the process, only $n$ and $\gamma$ are simulated directly, instead of the entire chain shown in Equation~(\ref{eq:ambe}). The simulation indicates that the $e^+e^-$ pair deposits its energy predominantly in a single cube. The experience from the IBD analysis suggests that the most efficient way to select the positron annihilation cube (AC) is to look for the cube with the most energy in the event. The normalised distribution of the energy found in the AC candidate is shown in Figure~\ref{fig:ambe_es_max}. The distribution contains two groups of events: with (dashed blue) and without (solid orange) $e^+e^-$ pair. The latter shows expectedly two Compton edges issued by the 4.44~MeV gamma from AmBe and the 2.2~MeV gamma from $n$-capture on hydrogen. It also contains proton recoil signals coming from the scattering with AmBe neutrons that are energetic enough, unlike the IBD, to trigger this physics process.     

The dashed blue signal distribution in Figure~\ref{fig:ambe_es_max} shows a sharp peak at 3.4 MeV of energy deposited by the $e^+e^-$ conversion pair. A small shoulder on the right-hand side of this peak is observed and corresponds to additional energy deposits from the annihilation gammas experiencing Compton scattering in the AC. The topological selection of calibration candidates of two back-to-back photons separated by several cubes (as performed with IBD signal candidates discussed in Ref.~\cite{reco_note}) removes most of this contribution.

\begin{figure}[ht!]
    \centering
    \includegraphics[width=0.475\textwidth]{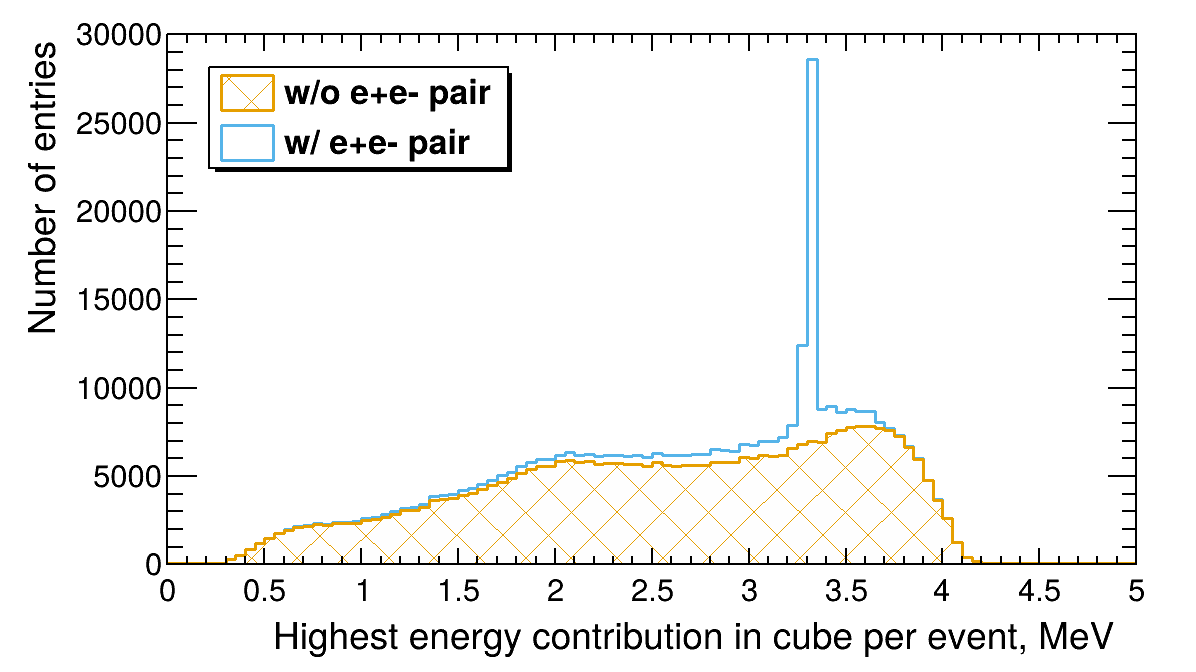}
    \includegraphics[width=0.475\textwidth]{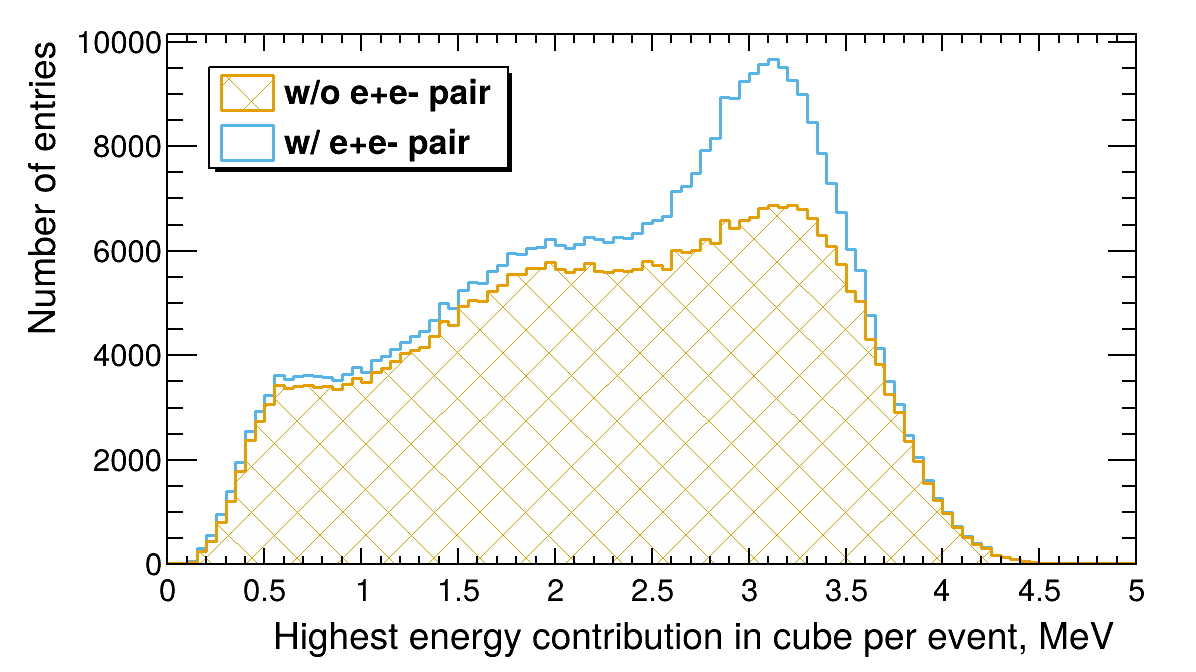}
    \caption{The stacked energy distribution of the cubes with the largest energy deposit in the events from the AmBe \textsc{Geant4} simulation (left) and ROSim output (right) with (empty blue) and without (orange diagonal cross-hatch) $e^+e^-$ pair conversion. Selected events have an empty envelope and 2 reconstructed annihilation gamma candidates. The smearing observed in the right plot corresponds to the emulation of the energy resolution of the detector ($\sim$15\%/$\sqrt{E}$).}
    \label{fig:ambe_s_and_b_g4_and_rosim}
\end{figure}

This \textsc{Geant4} simulation is processed with the SoLid reconstruction software Saffron2. Basic physics information about the event, such as the additional cubes in the envelope, the number of reconstructed annihilation gamma candidates, and the cosine of the angle between the annihilation gamma candidates, is built \footnote{The processing is done with the calibration based on the use of the horizontal muons and within the special module called the ReadOut Simulation or ROSim. The goal of this module is to make the \textsc{Geant4} output files compatible with the Saffron2 input format. ROSim transforms the \textsc{Geant4} energy deposits into the number of scintillation photons that are read out by the correspondent MPPCs. However, it skips the actual optical simulation and generation of the waveforms to optimise the computational cost of the processing. The detailed tuning of the ROSim is reported in Ref.~\cite{maja}.}.
Both signal and background events are preselected according to the geometric requirements : AC must be a unique cube in the envelope, the number of reconstructed annihilation gamma candidates must be 2, and they must be located in different hemispheres of the detector. 

Figure~\ref{fig:ambe_s_and_b_g4_and_rosim} shows the stacked signal (empty blue) and background (orange diagonal cross-hatch) distributions in the case of \textsc{Geant4} simulation (left) and ROSim (right). One can notice that the imposed geometrical requirements removed the annihilation gamma-knee energy deposits. The energy resolution of the detector results in a smearing of the measured distribution. The most probable values of both the signal and the background are found in the calibration region of interest.

\subsubsection{Calibration method}

Template models based on the \textsc{Geant4} energy distributions for the signal and background are used to interpret the AmBe calibration data following a method discussed in Section 3.4.2 in Ref.~\cite{roy_thesis}. The original model described in the previous section is first smeared event-by-event with a normal distribution. Namely, each cube energy is randomly varied following a standard distribution with $\sigma_E/E$ = $15\%/\sqrt{E}$ as indicated by $^{22}$Na calibration results~\cite{roy_thesis}. Each resulting distribution is further scaled by an absolute energy scale factor.

\begin{figure}[ht!]
    \centering
    \includegraphics[width=0.475\textwidth]{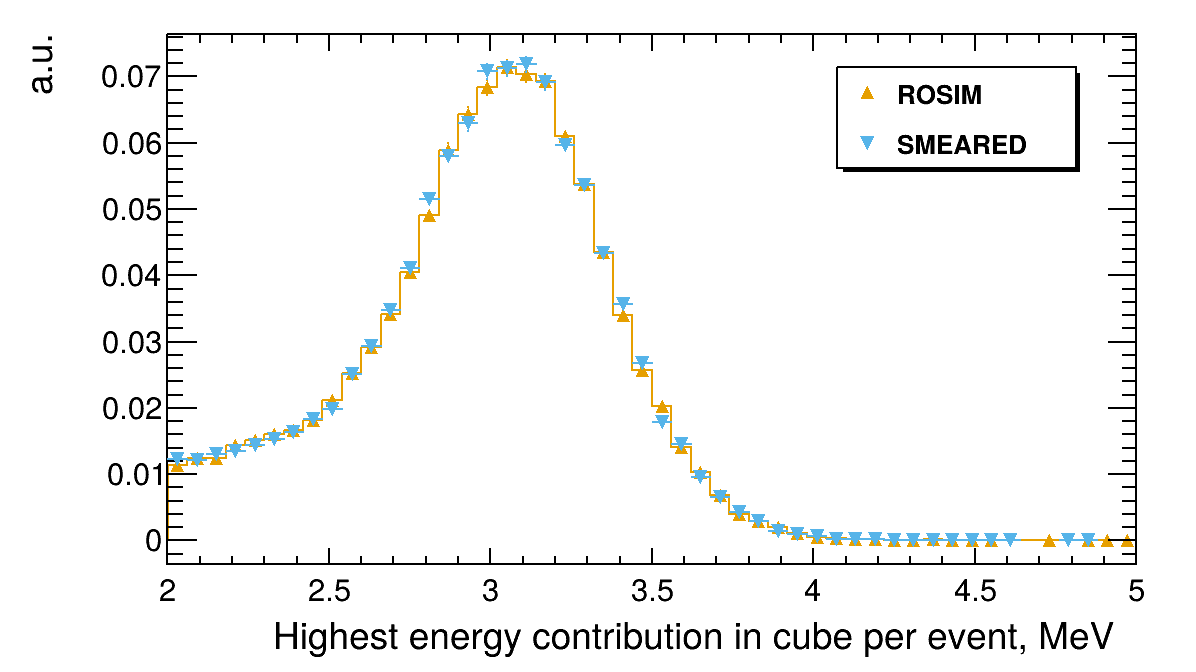}
    \includegraphics[width=0.475\textwidth]{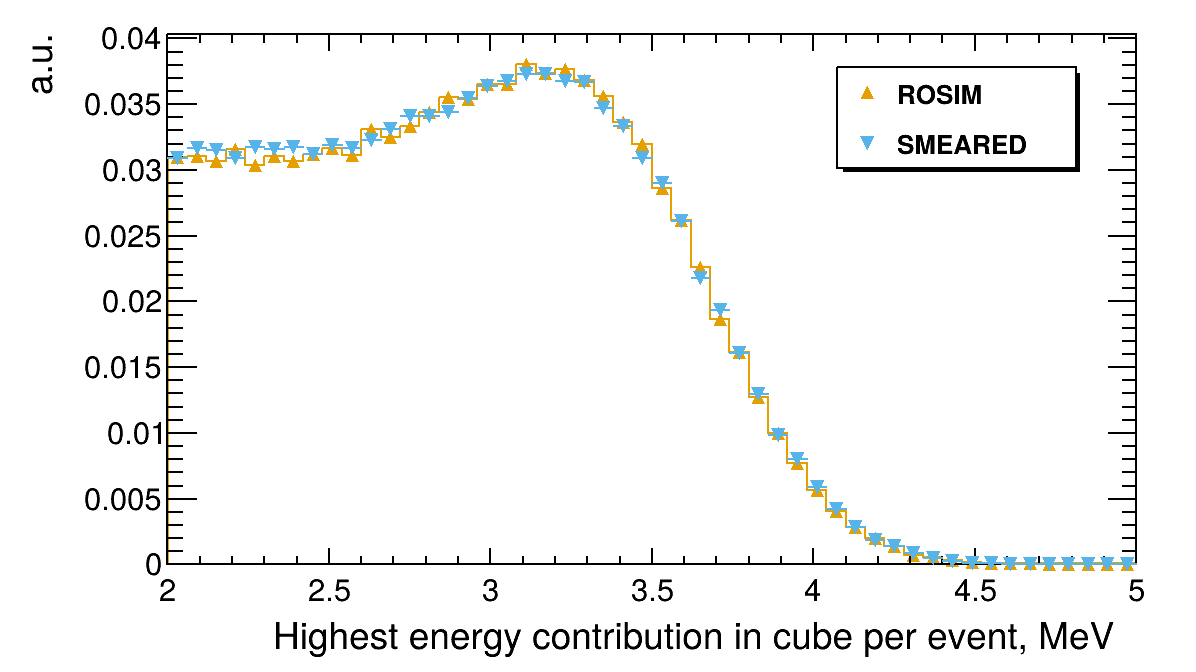}
    \caption{The comparison of the ROSim distribution with the \textsc{Geant4} template transformed with the best pair of parameters obtained from the $\chi^2$ scan.}
    \label{fig:ambe_rosim_sb_matching}
\end{figure}

\begin{figure}[ht!]
    \centering
    \includegraphics[width=0.8\textwidth]{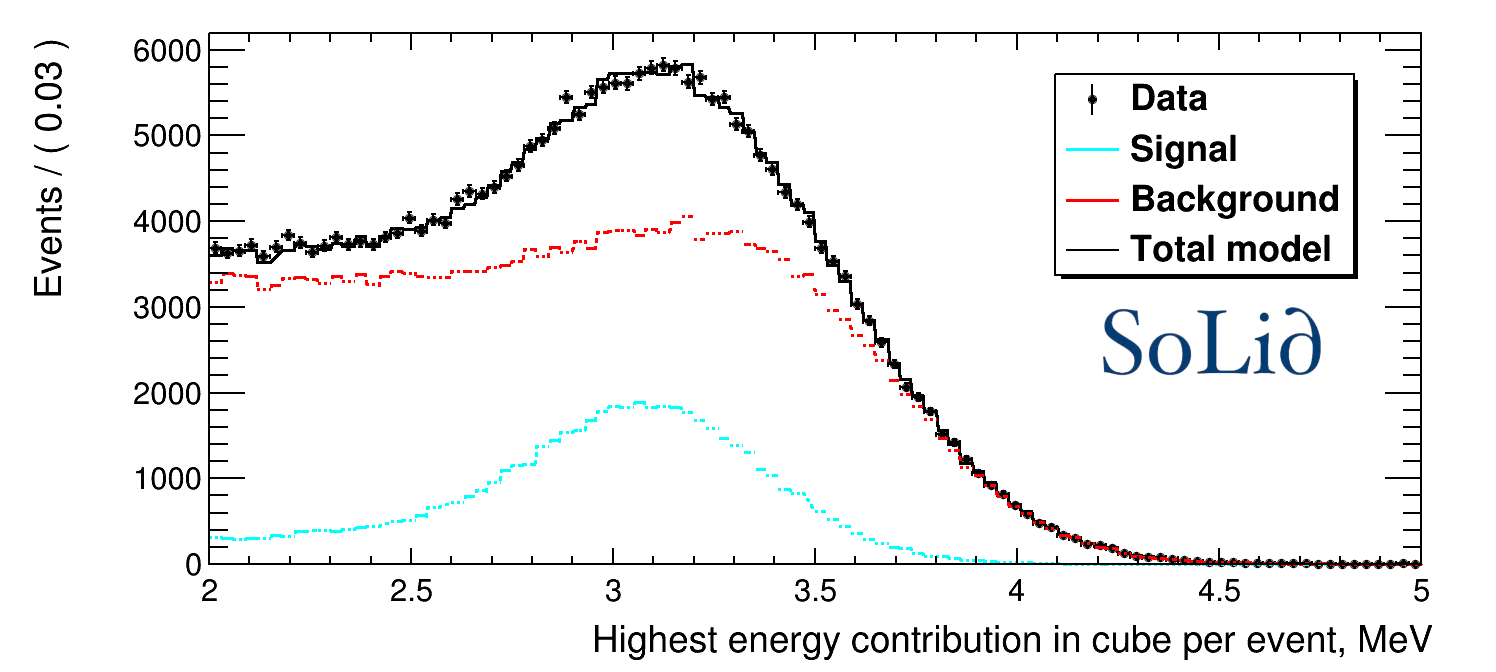}
    \caption{An example of the unbinned event-by-event ML fit performed on the AmBe ROSim sample. The background component is displayed in red, the signal component in blue, the total model in black, and the data are represented by the black points.}
    \label{fig:ambe_fit_example_rosim}
\end{figure}

The calibration proceeds via a scan of the parameter space for both the $\sigma$ and the absolute energy scale factor to find the best pair of parameters. An unbinned event-by-event Maximum Likelihood (ML) fit is performed with the ROOT RooFit toolkit software~\cite{Verkerke:2003ir}. For each pair, the transformed templates are converted to probability density functions, while the signal and background yields remain the free parameters of the fit. The negative logarithmic likelihood is used as a measurement of the goodness of fit. The method has been tested with ROSim data samples, separately for signal and background candidates first (see Figure~\ref{fig:ambe_rosim_sb_matching}) and then by combining them. The best fit is obtained for the pair of parameters (0.932, 14.9\%) and is reported in Figure~\ref{fig:ambe_fit_example_rosim}.

\subsubsection{Uncertainty budget}

\begin{figure}[ht!]
    \centering
    \includegraphics[width=0.475\textwidth]{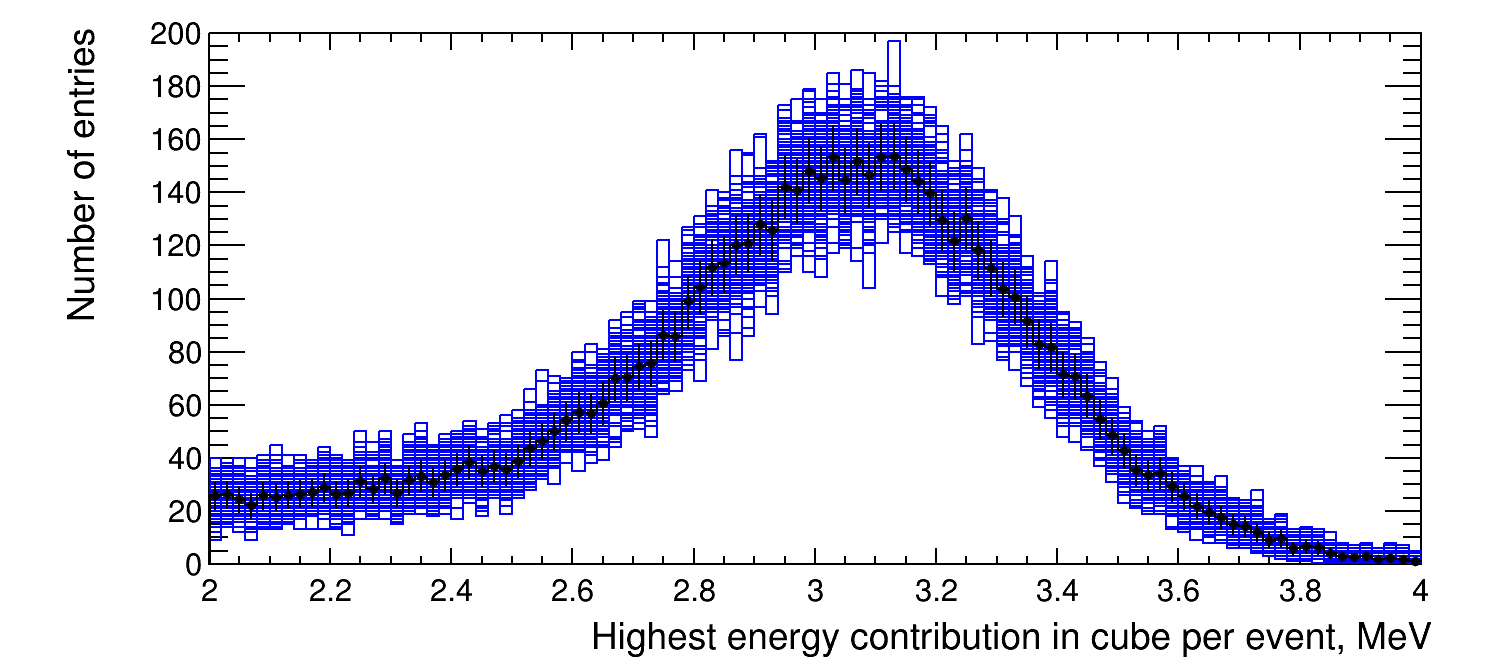}
    \includegraphics[width=0.475\textwidth]{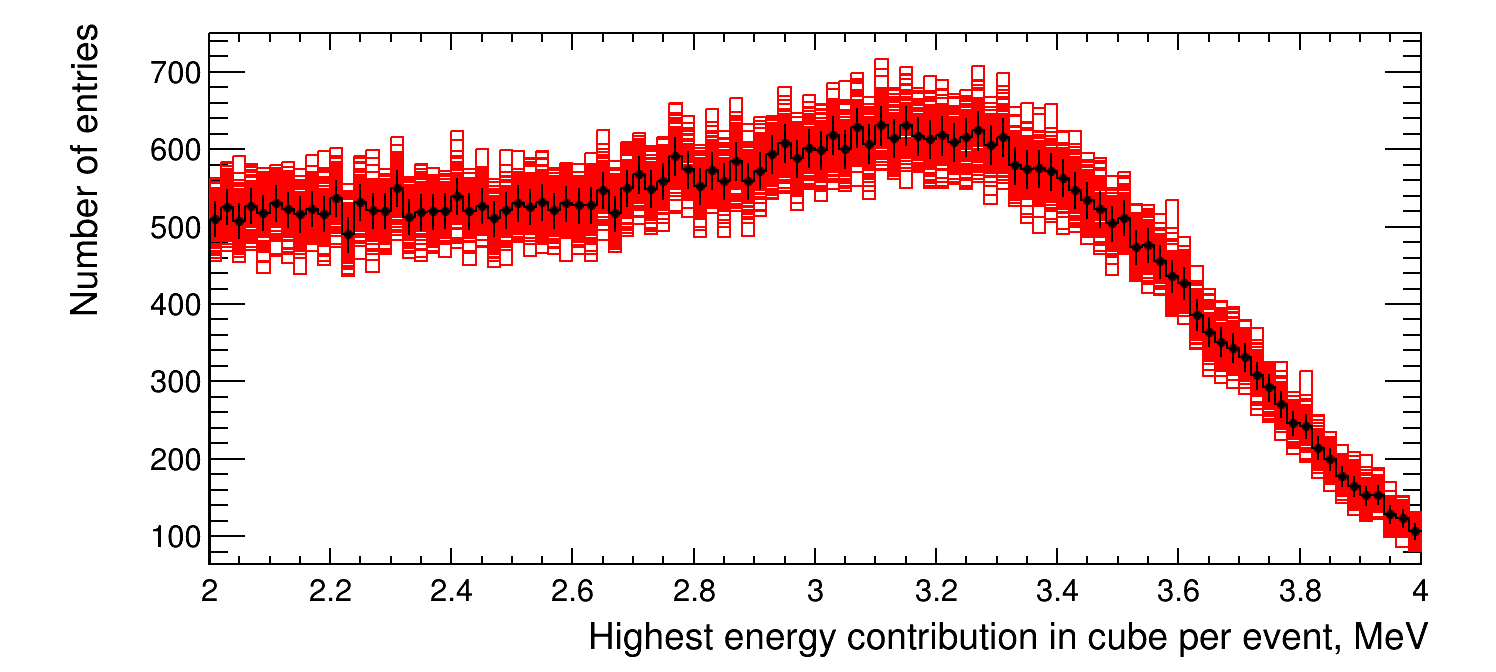}
    \caption{Sampled \textsc{Geant4} templates used for the statistical precision studies for signal (left) and background (right). The black data points indicate the baseline template.}
    \label{fig:stat_unc}
\end{figure}

\begin{figure}[ht!]
    \centering
    \includegraphics[width=0.775\textwidth]{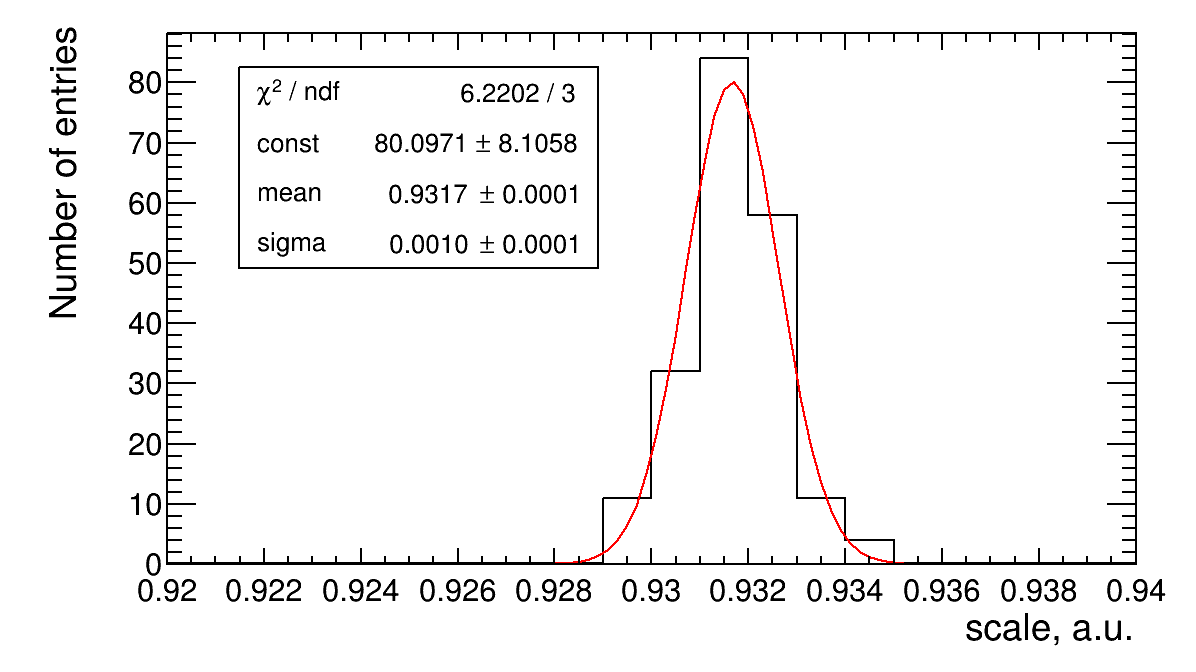}
    \vspace{-0.8cm}
    \caption{The scale factor values distribution obtained from the ROSim data fits with the sampled signal and background templates. The Gaussian fits shows the statistical precision at a sub-percent level.}
    \label{fig:scale_unc_stat}
\end{figure}

The statistical component of the uncertainty of the method is addressed by means of pseudoexperiments. The \textsc{Geant4} templates featuring the best pair of parameters obtained from the ROSim fit are taken as a baseline and used to generate the pseudoexperiments within the RooFit package. Figure~\ref{fig:stat_unc} shows the superimposed distribution for all pseudoexperiments with the baseline template indicated by the black data points. Each generation of pseudoexperiments is used to repeat the fit and obtain a new value for the absolute energy scale factor. The distribution of the values obtained is shown in Figure~\ref{fig:scale_unc_stat}. The spread of the distribution evaluated with a Gaussian fit provides the statistical precision of the method: it is less than one percent. 

\begin{table}[ht!]
    \centering
    \caption{The fit results with the modified model and ROSim sample.}
    \begin{tabular}{cc|c|c|c|c}
    \hline
    & Modification level: & 0\% & 5\% & 10\% & 20\% \\
    \hline\hline
    
    \multirow{2}{*}{\small{Modified Model}} 
    & \small{Scale} & 93.16 & 93.24 & 93.35 & 93.52\\
    & \small{Resolution} & 14.45 & 14.10 & 13.96 & 13.90\\
    \hline
    \multirow{2}{*}{\small{Modified ROSim sample}} 
    & \small{Scale} & 93.16 & 92.99 & 92.87 & 92.63\\
    & \small{Resolution} & 14.45 & 15.35 & 16.02 & 17.45\\
    \hline
     \end{tabular}
     \label{tab:ambe_syst}
\end{table}

The systematic studies focus on the possibility of the presence of an additional background component, which would not have been simulated. In fact, different background-to-signal ratios are observed in ROSim and calibration data. Indeed, the two AmBe calibration campaigns performed during data collection used trigger settings that were not aligned with those used in IBD data-taking periods, and no careful evaluations were performed on possible additional backgrounds. Furthermore, the \textsc{Geant4} simulation does not take into account the neutron pile-up (only one neutron is generated per event). This could result in additional proton recoils that are not simulated.        

In order to evaluate how robust the determination of the absolute energy scale factor is to an additional background, the scan of the best pair of parameters is performed in the presence of a linearly-decreasing background component with different global contamination fractions (5, 10 or 20\%). The results are collected in Table~\ref{tab:ambe_syst}. The determination of the energy scale factor is stable even with the highest contamination, suggesting that the systematic budget is comparable to the statistical
uncertainty. In contrast, the energy resolution parameter strongly depends on the presence of this additional potential background. This cannot thus be used reliably to determine the intrinsic energy resolution of the SoLid detector.  

\subsubsection{Results}

\begin{figure}[ht!]
    \centering
    \includegraphics[width=1.\textwidth]{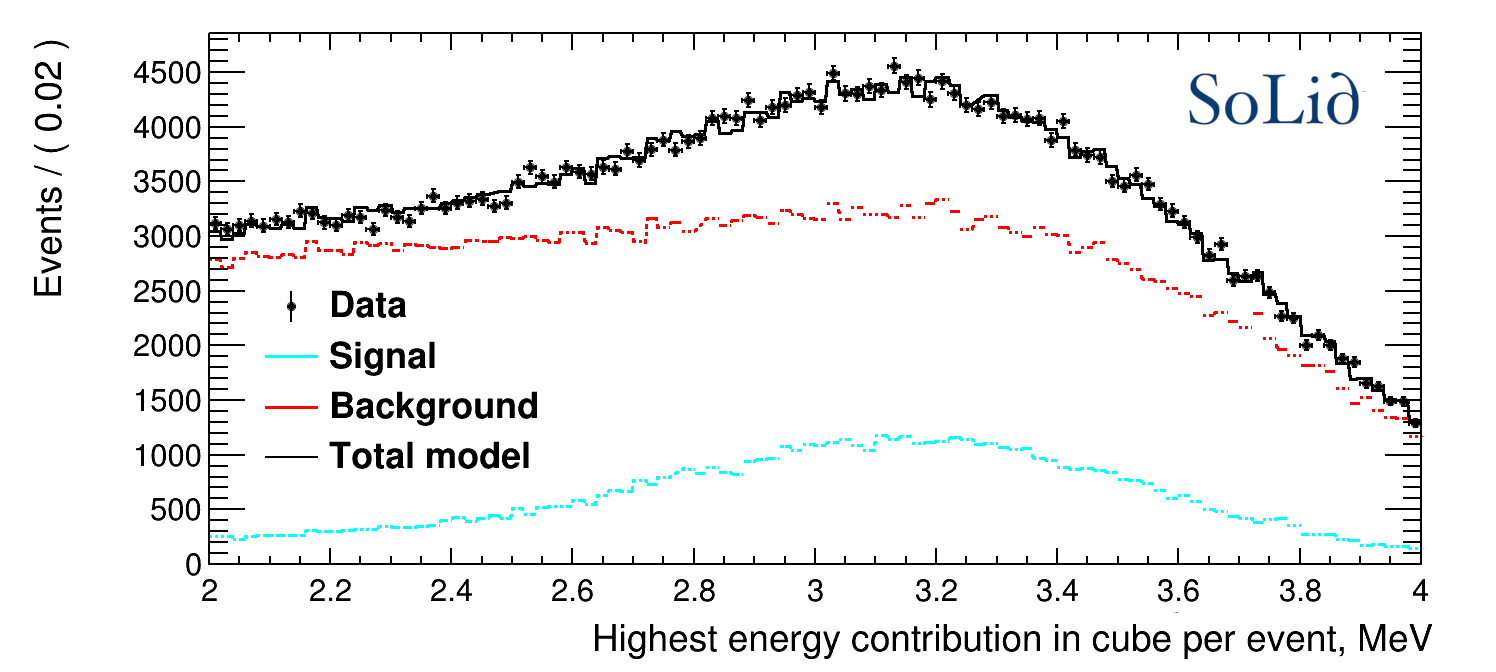}
    \caption{The result of the unbinned event-by-event ML fit performed on the AmBe sample.}
    \label{fig:ambe_fit_example_data}
\end{figure}

The same fit strategy used for the simulation is applied to the calibration data obtained in the 2020 summer calibration campaign. This campaign (as well as any other that involves the radioactive sources) in the SoLid experiment is performed with the CROSS (CalibRation On Site Solid) robotic hand. It is capable of inserting the source between the module and placing it in one of the nine predefined positions. The calibration data used in this paper was obtained with the source located between the last two modules at four different locations (central, lower left, lower right, and upper right). The total data-taking time was $\sim$ 77 hours. An additional requirement based on the low multiplicity of the cubes (less than 5) is applied (and by construction, an event must have at least 3 cubes). Indeed, events with low multiplicity only accommodate the AC and 1 or 2 cubes for each annihilation gamma. This requirement, therefore, mitigates the contributions of proton recoils in the case of events with multiple neutrons. The best-fit result is shown in Figure~\ref{fig:ambe_fit_example_data}. The value of the energy scale factor relative to the muon absolute calibration reads $0.9620 \pm 0.0014$.
\section{Crosscheck and validations}
\label{sec:crosschecks}

\begin{figure}[ht!]
    \centering
    \includegraphics[width=0.6\textwidth]{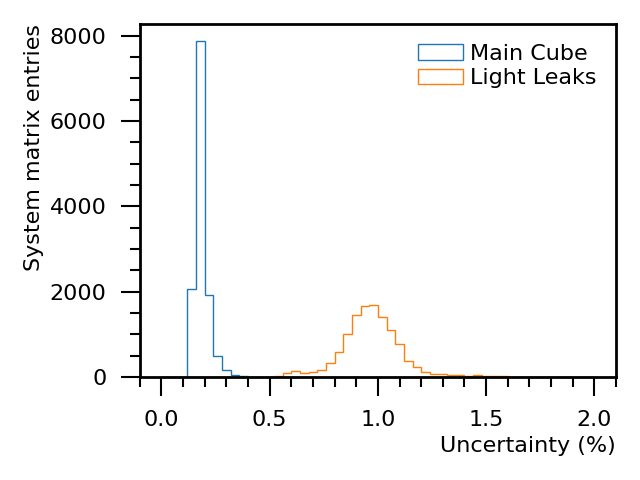}
    \caption{The distribution of the statistical uncertainty associated to the light sharing determination based on 10-days statistics of horizontal muons.}
    \label{fig:relative_muons_stats}
\end{figure}

Note that the results reported in this section use both relative and absolute calibration performed with the horizontal muons. To begin with, one of the calibration requirements was a statistical precision at the level of 1\%. To achieve it, the ROff data is gathered in 10-day periods. This sample size allows the precision of the method and the actual variations of the detector response to be aligned. Figure~\ref{fig:relative_muons_stats} shows an example of the statistical uncertainty distribution. The blue value corresponds to an uncertainty associated with one of the fibres in the main cube for each of the detection cells of the SoLid detector, whereas the orange distribution shows the same for one of the fibres impacted by the LL. 

Horizontal muons provide information on the time evolution of the detector response. The PVT is exposed to ageing, which decreases the performance of scintillating photon generation. Ageing is not homogeneous; hence, some of the fibres can be more exposed to it. Finally, the WLS fibres also undergo ageing, but the scale of the effect is much less significant, according to the manufacturer's information. All of these factors imply that the characteristics of light sharing evolve over time. 

\begin{figure}[ht!]
    \centering
    \includegraphics[width=7.25cm, height=6.5cm]{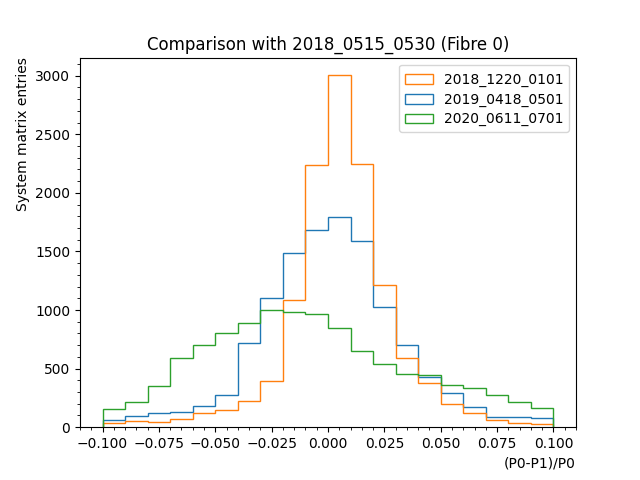}
    \includegraphics[width=6.75cm, height=6cm]{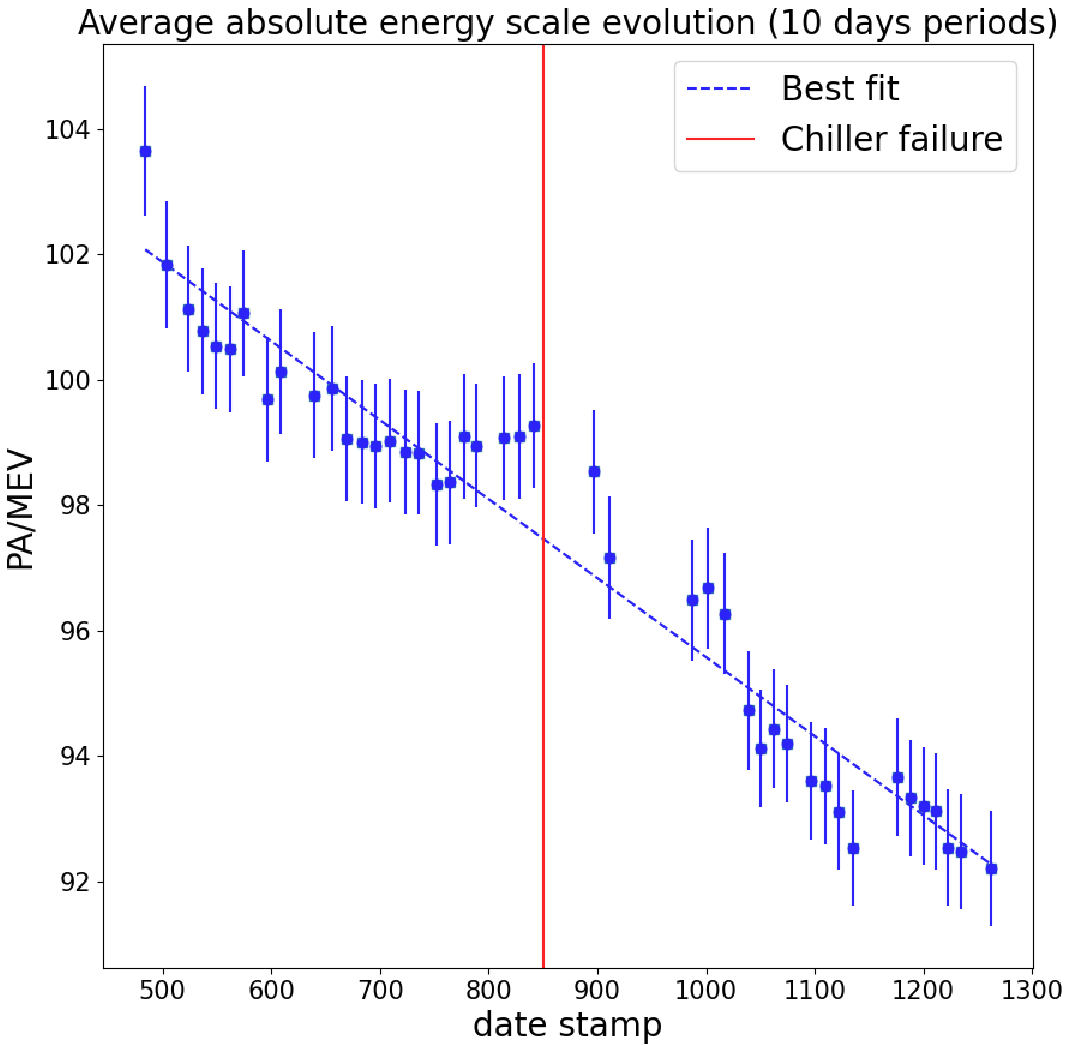}
    \caption{Left: the relative time evolution of the light sharing in a fibre of the main cube. The reference period is from the start of the detector operation in summer of 2018 while the others are from 6 (orange), 12 (blue) and 18 (green) months later. Right: the average absolute energy scale factor versus time. The dashed line shows the linear fit, that indicates the 4.5\% PVT ageing effect. The red solid line indicates the time stamp of the chiller failure.}
    \label{fig:light_sharing_time_evol}
\end{figure}

To demonstrate it, we take one of the vertical fibres in the main cube of each detection cell. The fraction of light ($a_{ij}$ value) received by this fibre during the reference period is compared with the fraction received by the same fibre in the following periods. The $x$ - axis in Figure~\ref{fig:light_sharing_time_evol}-left shows the evolution of the spread of the $a_{ij}$ value by computing the fraction $(a_{ij}^{P_0} - a_{ij}^{P_i}) / a_{ij}^{P_0}$. Where $P_0$ corresponds to the period from the very beginning of data-taking in summer 2018, and $P_i$ are the periods 6, 12, and 18 months later, respectively. The mean value of the distribution that shifts towards the negative one is a clear measurement of the ageing of the detector. Simultaneously, the growing width of the distribution shows that ageing impacts the fibres differently; hence the importance of the frequent relative calibration of the detector. The same behaviour is obtained for the four main fibre distributions.

\begin{figure}[ht!]
    \centering
    \includegraphics[width=0.475\textwidth]{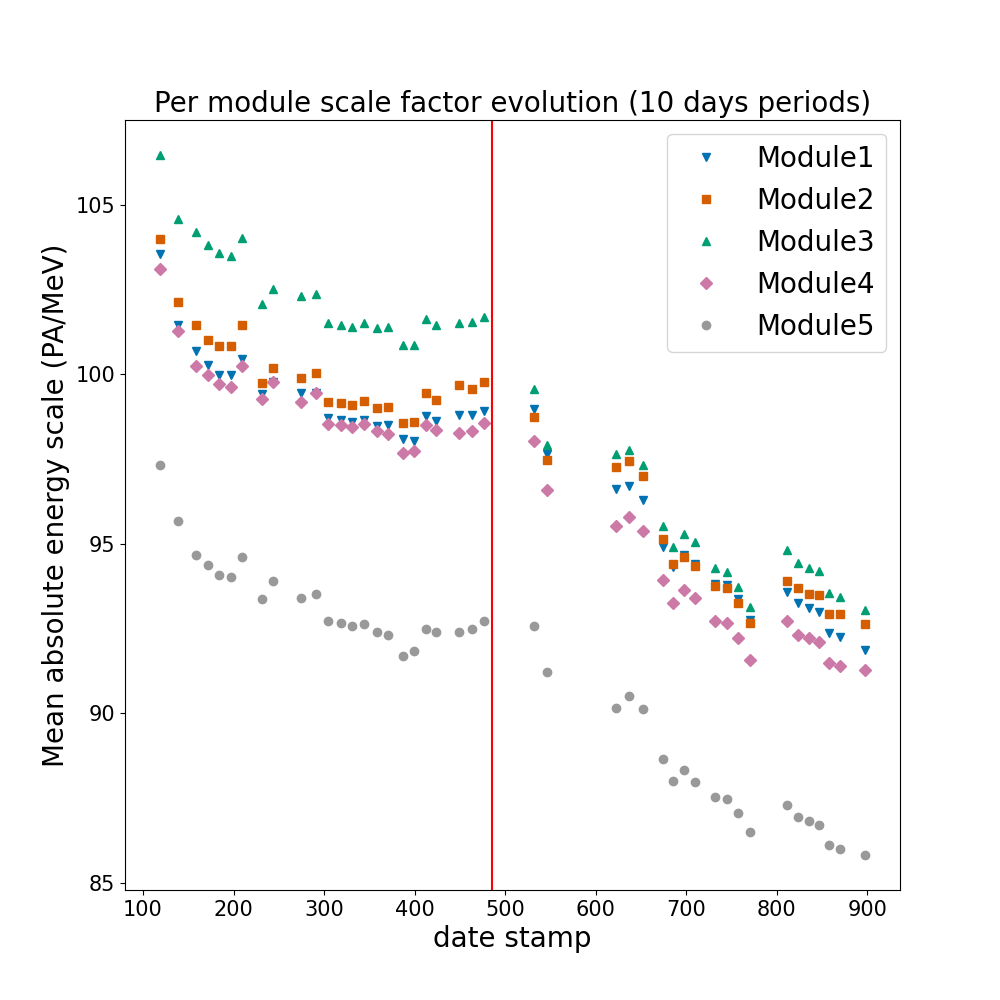}
    \includegraphics[width=0.475\textwidth]{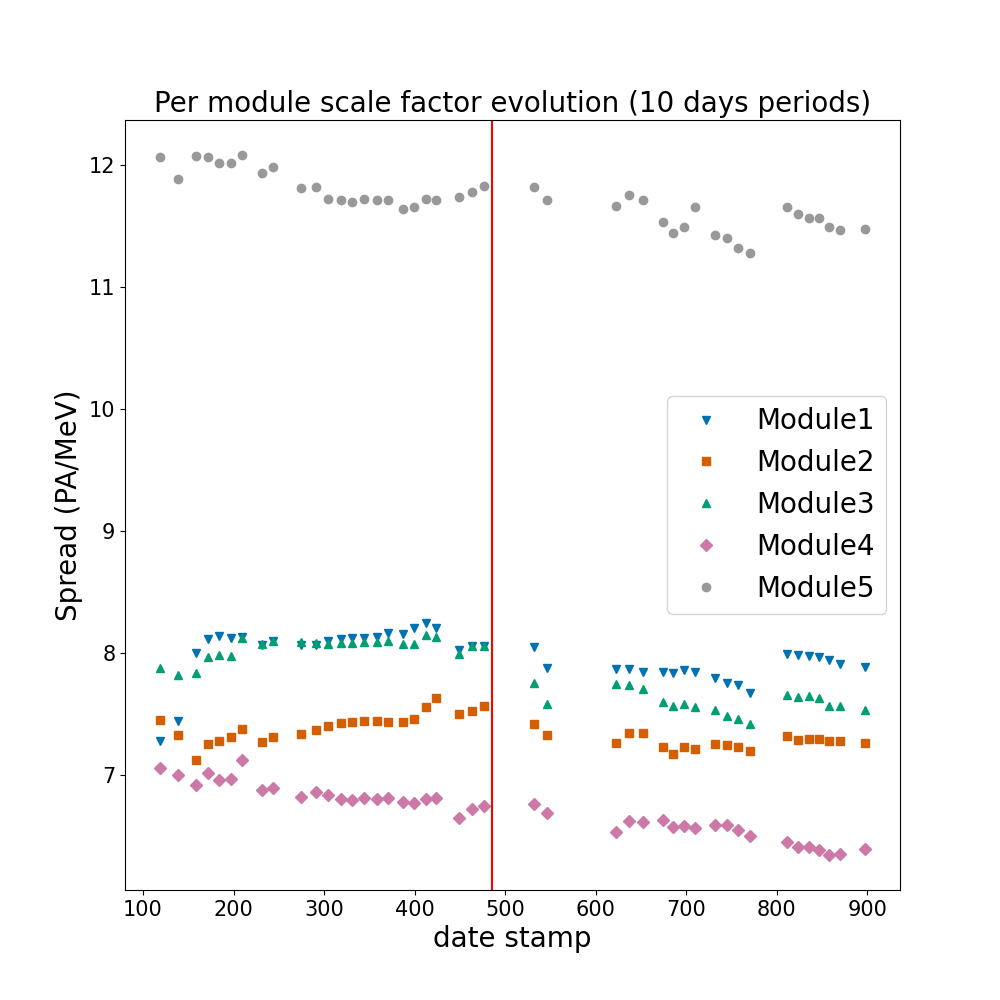}
    \caption{The average absolute energy scale factor (left) and the spread (right) per module for the 10 days statistics of horizontal muons versus time. The red solid line indicated the time stamp of the chiller failure.}
    \label{fig:ly_evolution_per_module}
\end{figure}

The absolute energy scale can be used as an alternative approach to track the evolution of the detector response. The plot on the right of Figure~\ref{fig:light_sharing_time_evol} shows the evolution of the absolute energy scale factor. This is yet another demonstration of the ageing of PVT. The average effect is around 4 - 5\% per year, which is slightly higher than expectation~\cite{yellowing}. It could be explained by a chiller failure that occurred in May 2019, when the temperature in the detector container increased to 50$^\circ$C instead of the usual 12$^\circ$C in standard operation. The incident is believed to have caused a degradation in the performance of the PVT. The final cross-check is done with a higher granularity. It explores the evolution of the response on the per-module level. The result is presented in Figure~\ref{fig:ly_evolution_per_module}. The last detector module concentrates most of the dead fibres; thus, the average absolute energy scale in the module is expected to be lower, while it is determined with slightly worse precision. This effect is mitigated if planes beyond plane 45, where most of the dead fibres reside, are discarded. Without these planes, the performance of module 5 matches the performance of other modules. 

\subsection{Cosmogenic background determination}
\label{subsec:b12}

\begin{figure}[ht!]
    \centering
    \includegraphics[width=0.8\textwidth]{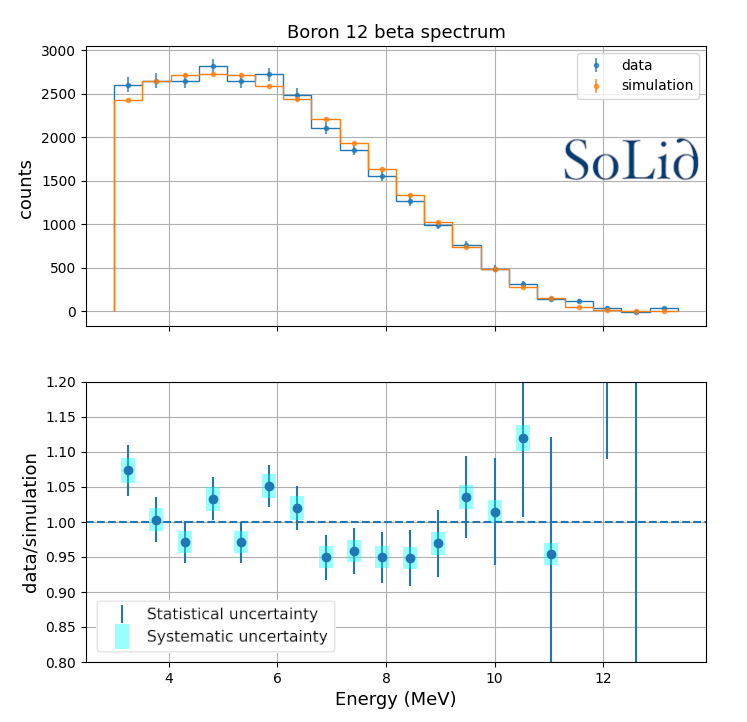}
    \caption{The comparison of the $^{12}$B energy spectrum from ROff (blue) and \textsc{Geant4} (orange). Bottom: relative ratio between the two with the statistical (blue bars) and systematic (cyan rectangles) uncertainties.}
    \label{fig:b12_rosim_vs_g4}
\end{figure}

Standard background candles can be used to verify the combined calibration and reconstruction tools. The energy distribution is measured and can be compared with the \textsc{Geant4} prediction. It has been done for cosmogenic muon-induced backgrounds issued from the excitation of the $^{12}$C nuclei that the PVT comprises: 

\begin{equation}
\begin{gathered}
        \ce{\mu- + ^{12}C -> ^{12}B + \bar{\nu}_{\mu}} \\
        \ce{^{12}B -> ^{12}C + \bar{\nu}_{e} + e-}. \\
\end{gathered}
\end{equation}

\noindent $^{12}$B is a well-known radioactive $\beta$ source. The half-life of the isotope is short ($\sim$ 20.2 ms), and the end-point energy of subsequent $\beta$ decay is $\sim$ 13.4~MeV. Therefore, the energy distribution of these electrons completely covers the IBD energy range of interest. The selection of event candidates is based on the identification of muon tracks that stop inside the detector. Then an electromagnetic cluster, close in space and time to the reconstructed muon, is searched. To suppress the background, the energy of the cluster is required to be above 3~MeV. Furthermore, events with additional muon candidates within the 200~$\mu$s time window are removed. It suppresses spallation neutrons that can mimic an electromagnetic cluster. The yield of $^{12}$B is estimated from a fit of the time distance between the muon and the electromagnetic cluster. Finally, the energy spectrum reported in Figure~\ref{fig:b12_rosim_vs_g4} is determined after a statistical subtraction of the background using the sPlot~\cite{splot} technique. The comparison between the measured spectrum and the reference from \textsc{Geant4} shows an agreement better than 5\% throughout the energy range. These results validate both relative and absolute calibration with horizontal muons. This standard candle background could not be observed in the SoLid data prior to the introduction of the muon calibration. The interested reader can find a comprehensive description of this analysis in Ref.~\cite{savitri}.

\subsection{The linearity curve}

\begin{figure}[ht!]
    \centering
    \includegraphics[width=0.8\textwidth]{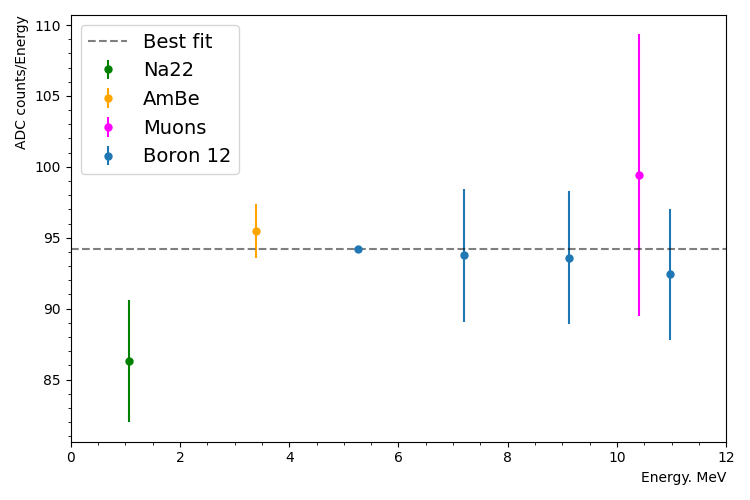}
    \caption{The linearity of the SoLid detector energy response measured with 4 calibration sources that covers the typical reactor antineutrino energy range. $^{22}$Na point is shown with the statistical uncertainty only.}
    \label{fig:linearity}
\end{figure}

It is interesting to gather the different calibration results obtained with the SoLid detector in a single linearity plot. The plot shows the calibration values obtained from the horizontal cosmic muons, the cosmogenic background source $^{12}$B, and the AmBe radioactive source measurements covered in this paper. It is complemented with an additional point provided by another radioactive source $^{22}$Na, which was previously used by the collaboration and is discussed extensively in Ref.~\cite{Henaff:2021paw}. A coherent picture emerges from this figure and provides confidence in the standard calibration tools of the detector response used for the physics measurements discussed in Ref.~\cite{osc_note}.
\section{Conclusions}
\label{sec:conclusions}

This paper describes the relative and absolute energy calibration of the positron signal in the SoLid detector. An accurate reconstruction of the reactor antineutrino energy is of utmost importance for both oscillation and ``5~MeV bump'' analyses. The relative calibration of the detector cells is performed at the percent level thanks to the large samples of horizontal cosmic muons recorded by the SoLid detector. Subsequently, the same horizontal muon sample is used to provide the absolute energy scale factor with a systematic uncertainty at the level of 10\%. A novel alternative method that relies on the creation of electron-positron pairs from the conversion of AmBe 4.4 MeV $\gamma$ into the detector matter is introduced and allows for an absolute energy calibration at the percent level. The methods and procedures have been further validated by the successful search for a cosmogenic background candle $^{12}$B and its subsequent measurements.

\acknowledgments
This work was supported by the following funding agencies: Agence Nationale de la Recherche grant ANR-16CE31001803, Institut Carnot Mines, CNRS/IN2P3, and Region Pays de Loire, France; FWO-Vlaanderen and the Vlaamse Herculesstichting, Belgium; The UK groups acknowledge the support of the Science \& Technology Facilities Council (STFC), United Kingdom; We are grateful for the early support given by the subdepartment of Particle Physics at Oxford and High Energy Physics at Imperial College London. The authors also thank our colleagues, the administrative and technical staff of the SCK CEN, for their invaluable support for this project. Individuals have received support from FWO-Vlaanderen and the Belgian Federal Science Policy Office (BelSpo) under the IUAP network programme; the STFC Rutherford Fellowship programme and the European Research Council under the European Union Horizon 2020 Programme (H2020-CoG)/ERC Grant Agreement n. 682474; Merton College Oxford.

% Bibliography

%% [A] Recommended: using JHEP.bst file
\bibliographystyle{JHEP}
\bibliography{biblio.bib}

%% or
%% [B] Manual formatting (see below)
%% (i) We suggest to always provide author, title and journal data or doi:
%% in short all the informations that clearly identify a document.
%% (ii) please avoid comments such as "For a review'', "For some examples",
%% "and references therein" or move them in the text. In general, please leave only references in the bibliography and move all
%% accessory text in footnotes.
%% (iii) Also, please have only one work for each \bibitem.

\end{document}